# Multi-objective Bayesian optimization reveals HiPIMS deposition rates exceeding direct current sputtering

Nathan Rodkey[1]*, Felix Thelen[2], Alfred Ludwig[2,3], Sebastian Siol[1]*

[1] Empa, Swiss Federal Laboratories for Materials Science and Technology, Laboratory for Surface Science and Coating Technologies, Überlandstrasse 129, 8600 Dübendorf, Switzerland

[2] Materials Discovery and Interfaces, Institute for Materials, Ruhr University Bochum, 44801 Bochum, Germany

[3] Center for Interface-Dominated High-Performance Materials (ZGH), Research Center Future Energy Materials and Systems (RC FEMS), Ruhr University Bochum, 44801 Bochum, Germany

∗Corresponding authors: nathan.rodkey@empa.ch, sebastian.siol@empa.ch

## Abstract

*Pulsed sputter processes are essential for a wide range of applications from hard coatings to microelectronics. High power impulse magnetron sputtering (HiPIMS) offers unique possibilities for process design due to its high ionized flux fraction. Because of the high peak-current densities involved, correspondingly high plasma densities generate many ions. This enables tuning of kinetic energies by applying accelerating bias potentials. However, some HiPIMS processes have low deposition rates due to the back-attraction these ions experience with the negatively charged sputter target. Recent work in HiPIMS has targeted this problem, exploring complex pulsing schemes such as bipolar, multi-pulse, or chopped HiPIMS. These pulsing schemes introduce another problem: process complexity. As such, the process space, even with coarse intervals quickly blows up to 1E5–1E20 possible process combinations. This calls for efficient and intelligent sampling, which can be enabled by machine-learning algorithms such as Bayesian optimization. In this work we use multi-objective Bayesian optimization (MOBO) to efficiently sample the process space of pulse-packets, targeting high ionization flux fractions and deposition rates. Over 2,000 distinct processes were sampled during this exploration. Surprisingly, we found that many HiPIMS processes, including single pulse, can achieve deposition rates higher than DC references. In fact, at a working pressure of 1.0 Pa and moderate ionization rates of 5–20% we found that ms-scale pulses were ideal. We hypothesize these processes maximize gas rarefaction, creating a low-collision channel between the target and the substrate. These results showcase previously unreported behavior of pulsed sputter processes and challenge a long-held trade-off between film quality and speed in thin-film manufacturing.*

## 1. Introduction

High-power impulse magnetron sputtering (HiPIMS) applies short, µs-scale voltage pulses to a target, driving peak current densities orders of magnitude above conventional sputtering and thereby ionizing a large fraction of the sputtered material.[1,2] The resulting ion-dominated deposition flux densifies films, improves adhesion and conformality, and allows for substrate biasing to tune the energy of the arriving ions.[2,3] However, the same electric fields that ionize the sputtered material also pull a large fraction of it back to the target. This effect is known as back-attraction and significantly reduces the deposition rate. In fact for several decades, the consensus in the community was, that at constant average power, the deposition rate of HiPIMS falls well below that of conventional DC sputtering.[4–6]

Despite these reduced deposition rates, HiPIMS has seen substantial use for hard and protective coatings, diffusion barriers, and conformal metal coatings.[2,7,8] Recently, substantial work has been done to show that HiPIMS can also target defect-sensitive applications, including piezo- and ferro-electric devices[9–11] and transparent conducting oxides.[12] These applications were previously thought inaccessible to HiPIMS due to strong ion bombardment/implantation from high energy species, but a combination of gentler processing conditions and advanced accelerating bias schemes have enabled significant advances, such that material quality rivals those of standard processing techniques.[9,11,13]

However, industry may remain reluctant to adopt these techniques due to concerns over reduced deposition rates.

How much deposition rate and ionized flux fraction a pulsed process trades depends on how the power is pulsed. Mid-frequency pulsed-DC switches in the 10–200 kHz range at duty cycles typically above 80 %,[14,15] but its ionized flux fraction is similar to DC and on the order of a few percent only.[3] HiPIMS concentrates the same average power into pulses tens to hundreds of microseconds long at repetition frequencies of 50 Hz–10kHz and duty cycles below 10 %,[2,16] reaching ionized flux fractions of tens of percent and up to about 70 %.[1,3] Reported HiPIMS deposition rates routinely reach only 70–85 % when compared to their DC-sputtering equivalent, with part of the deficit being the sub-linear sputter yield at the higher voltage HiPIMS needs.[5]

A significant cause of the decreased deposition rate found in pulsed sputter processes, and in particular in HiPIMS, has to do with the high ion concentration. The negative voltage applied to the sputter target back-attracts these ions. This back-attraction falls when pulses are shortened,[17] when the pressure is lowered[18] and when the magnetic field is weakened.[19,20] Multi-pulse or chopped HiPIMS exploits the effect of shortened pulse-widths reducing back-attraction. Splitting one long pulse into a packet of shorter pulses separated by short gaps can raise the deposition rate substantially at comparable voltage and peak current, by as much as 45–150 % for Ti,[21,22] which has been attributed to gas rarefaction and to relaxed ion trapping between sub-pulses. This happens in part because back-attraction acts only while the pulse is on and vanishes in the afterglow, so a pulse packet, with an afterglow in between each of its sub-pulses, harvests ions that a single long pulse would have lost.[17] Consistent with this, shortening a pulse at fixed peak current can raise the deposition rate by about 45 % at constant ionized flux fraction.[23]

A second, less examined consequence of pulsing is that the discharge rarefies its own working gas. The sputtered flux expels gas from in front of the target by about 50 % at high peak currents, and the gas recovers only over hundreds of microseconds to milliseconds.[24–27] Where the gas is thin, ionization happens farther from the target, and an ion made there is better placed to escape the sputter target's electric field.[28] This coupling between gas dynamics and back-attraction can be exploited with tailored pulse packets to optimize the deposition rate, ionized flux fraction and even kinetic energy distribution for a given application.

However, a big obstacle to designing such processes is the size of the parameter space. A packet has a pulse count, a per-pulse width, a gap, and a repetition frequency. That can explode to more than $10^5$ recipes for moderate bounds and relatively coarse step-sizes on those parameters (**Table S1**). This estimate only considers symmetric pulses, each pulse in the packet having the same width and spacing. When moving to variable pulse packets the space can easily reach beyond $10^{20}$. As a result, many works publish results on ionization and deposition-rate interactions by varying just one parameter at a time on a few base discharges. The most complete constant-power dataset published to date, 36 copper discharges at 500 W, gives a strongly convex rate–ionization relation,[29] and the two chopped-pulse comparisons nearest to our experimental design disagree about whether pulse chopping is beneficial at all. *Hnilica et al.* find packets exceeding conventional DC sputter rates at equal average power only at 2.7 Pa and an ionized fraction near 20 %,[30] while *Farahani et al.* find a plain 20 µs pulse beating their own best multi-pulse process by 25 % on deposition rate.[31]

As such, there is a serious lack of data to better understand the interplay between deposition and ionization rates and mapping the boundary of such a space needs both significant and intelligent sampling. Machine-learning (ML) techniques such as closed-loop Bayesian optimization have huge promise for this kind of sampling. Despite this, ML studies are underutilized in thin film deposition with few effective reports.[32–35] Particularly in HiPIMS where we recently showed a proof of principle study, showing how powerful machine-learning algorithms can be when combined with complex HiPIMS pulsing schemes.[36]

Because both high deposition rates and high ionized flux fractions are important factors for the industrialization of HiPIMS processes, we present a multi-objective Bayesian optimization (MOBO) algorithm, which steers each measurement towards resolving a high ionization flux fraction and high deposition-rate front (see **Figure 1**).[37] As a result we establish the Pareto-front of these two critical parameters in HiPIMS process development. This is done for Ti, with each campaign bound by pulse count, width, frequency, and spacing. They are performed at different pressures and powers, resulting in >2,000 distinct processes and measurements, with a DC reference measured at each pressure and power. We find HiPIMS processes, including many single pulse processes, that surpass DC deposition rates significantly. Additionally, ion and neutral deposition rates appear to be dominated by their interactions with pressure, and their deposition rates approach a linear Pareto front when normalized for power. We hypothesize this has to do with the maximal achievable rarefaction of the process gas, which sets a lower floor for the effective back-attraction a process can reach. This is supported by the strong correlation Pareto front processes have with on-time, which suggests that optimal process conditions are for single pulses and not packets. In fact, we measured near identical ionization flux fractions and deposition rates between equivalent single pulses and pulse packets. As such, we propose design principles that rely on high initial peak currents to drive rarefaction, followed by a long shallow pulse that pushes ions and neutrals through the rarefied channel. These design principles work for maximizing deposition rates while maintaining modest ionization flux fractions of 5–25%. For higher ion counts we find that multi-pulse approaches are needed to co-optimize deposition rates and ionized flux-fractions. We expect the trends observed for Ti to translate to many other material systems.

## 2. Results & Discussion

Several thousand unique HiPIMS processes were evaluated, collected using a multi-objective Bayesian optimization (MOBO) function in a closed loop. This is shown schematically in **Figure 1** and described in detail in the *Experimental Methods* section. The closed-feedback loop starts with an initial random sample of processes, which it uses to fit a Gaussian Process Regressor (GPR) model to both the ionization and deposition rates. These surrogate models are then used to evaluate a Pareto front. An acquisition function then compares several potential processes based on their expected hypervolume gain. This next process is then executed, measuring its ion and deposition rates using a modified QCM, detailed in the *Experimental Methods*. The results from this process feed back into the surrogate models so that the process can repeat.

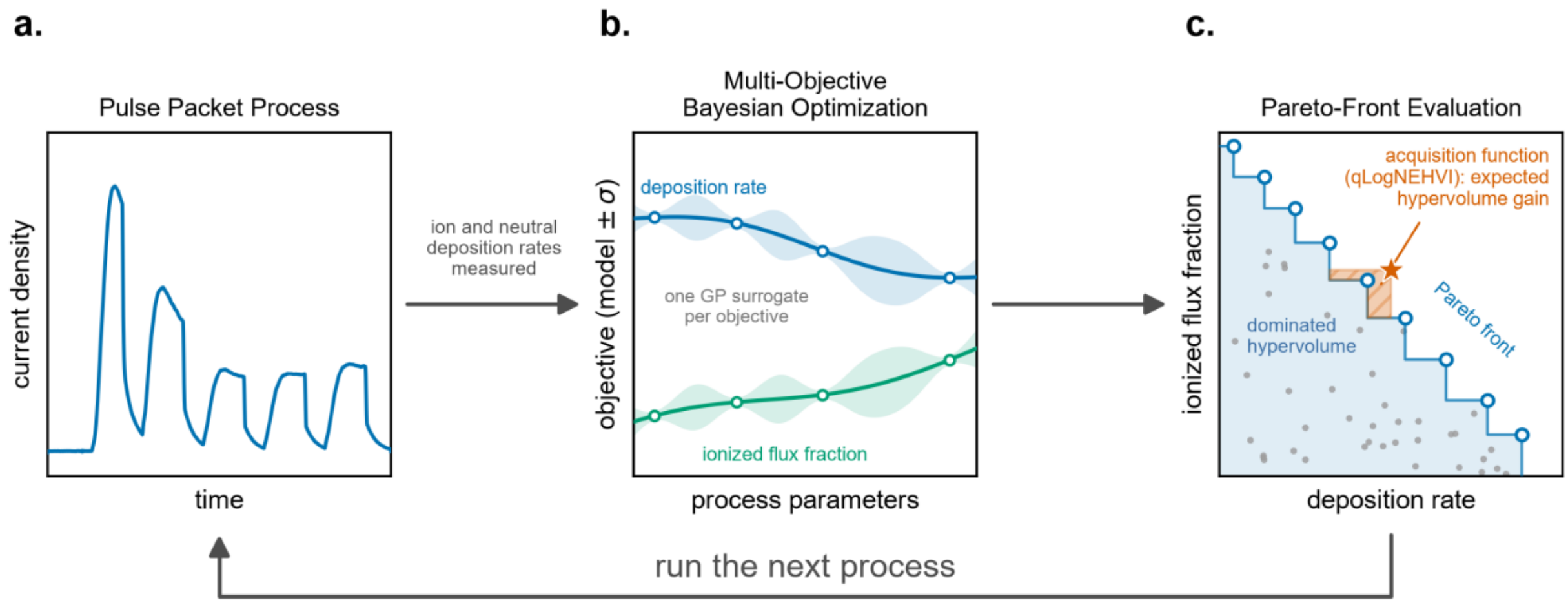

**Figure 1.** A schematic of the MOBO feedback-loop, which starts with an initial random sampling. **a.** The process that scored highest in the acquisition function is run next, measuring both its ionization and deposition rates. These results feed back into the surrogate models and the loop repeats. **b.** A GPR surrogate model fits independently on both the ionization and deposition rates. **c.** Using the surrogate models, a Pareto front is evaluated, and processes are graded by their expected hypervolume gain. This acquisition function uses qLogNEHVI (q + log + noisy + expected hypervolume improvement) which is described in detail in the *Experimental Methods*.

## 2.1 Pressure dependence and DC deposition-rate references

**Figure 2a** displays three deposition campaigns, using the MOBO workflow, at working pressures of 0.5, 1.0, and 2.0 Pa in a power-controlled mode of 2.2 $Wcm^{-2}$, with the DC deposition rate reference marked in a dashed black line. Each validated front maps the highest ionization flux fraction at a given deposition rate and is collected using a dual-objective acquisition function, described in *Experimental methods*. This is traditionally referred to as a *Pareto Front* in the ML community.

For several decades it has been assumed that pulsed sputtering at reduced duty cycles inevitably results in a reduction of deposition rate.[4–6] Strikingly, most of the optimized HiPIMS processes studied in this work surpass those of a DC reference. This is shown in **Figure 2b**, where all Pareto fronts of the 2.2 and 4.4 $Wcm^{-2}$ campaigns are shown. DC reference deposition rates are shown in **Figure S1**. At 2.2 $Wcm^{-2}$ the DC reference only outperforms HiPIMS at ionization flux fractions above ~35 %. This is marked by the dashed 'DC break-even' line. The three campaigns are shown in detail in **Figure 2c–e** with processes colored by the number of HiPIMS pulses in the pulse-packet. At 0.5 Pa, single pulses dominate the front, but as working pressures increase, higher pulse counts start to dominate the front. Additional data collected at 4.4 $Wcm^{-2}$ is shown in **Figure S2**.

Such a strong deviation from the usually reported HiPIMS deposition rate penalty is only partly established in the literature and may be attributed to the low ionization flux fractions, but also the much greater process parameter space our data covers. Traditional HiPIMS aims at far higher ionized flux fractions, up to about 70 % at very high peak power densities,[1] and extrapolated to such fractions our fronts fall well below the DC rate, in line with deficits most of the literature reports. Between these extremes the community's data is thin, so it is often unclear where a given process sits in this deposition-rate landscape. The few systematic comparisons that exist vary the magnetic field or the geometry rather than the pulse pattern, and both move the measured ratio severalfold.[19,20] In addition, the ion and neutral fluxes spread differently on the way to the substrate, so the their measured ratio also depends on the working distance.[38,39]

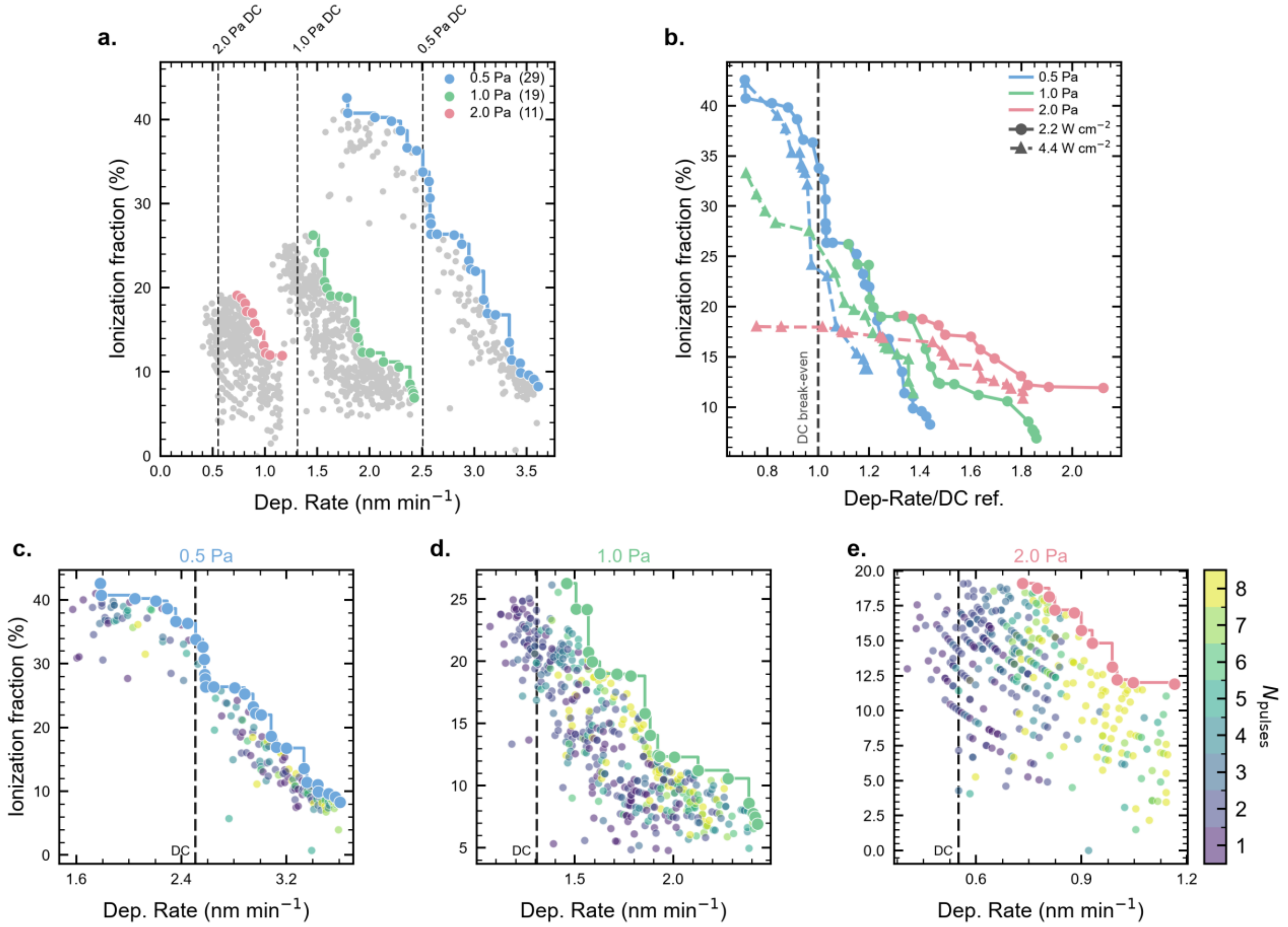

**Figure 2.** HiPIMS processes displayed alongside their DC deposition rate references. **a.** All three 2.2 Wcm$^{-2}$ campaigns show a significant impact of pressure on moving the Pareto front. All measured points are shown in grey, with validated Pareto-front points shown in color. A 2.2 Wcm$^{-2}$ DC reference deposition rate is marked as a black dashed line for each pressure and the number of processes on the front is shown in parentheses to the right of the working pressure. **b.** Deposition rate along each Pareto front, normalized to the DC reference deposition rate for equivalent power and pressure. This is shown for campaigns run at two different power densities, 2.2 and 4.4 Wcm$^{-2}$ marked by circles and triangles respectively. The DC break-even is marked by a dashed black line, with most HiPIMS processes surpassing this deposition rate. Only at ionization fractions greater than ~35 % do DC deposition rates start to pass those of HiPIMS. In **c–e.** The same three campaigns, colored by pulse count. Equivalent campaigns deposited at 4.4 Wcm$^{-2}$ are shown in **Figure S2**.

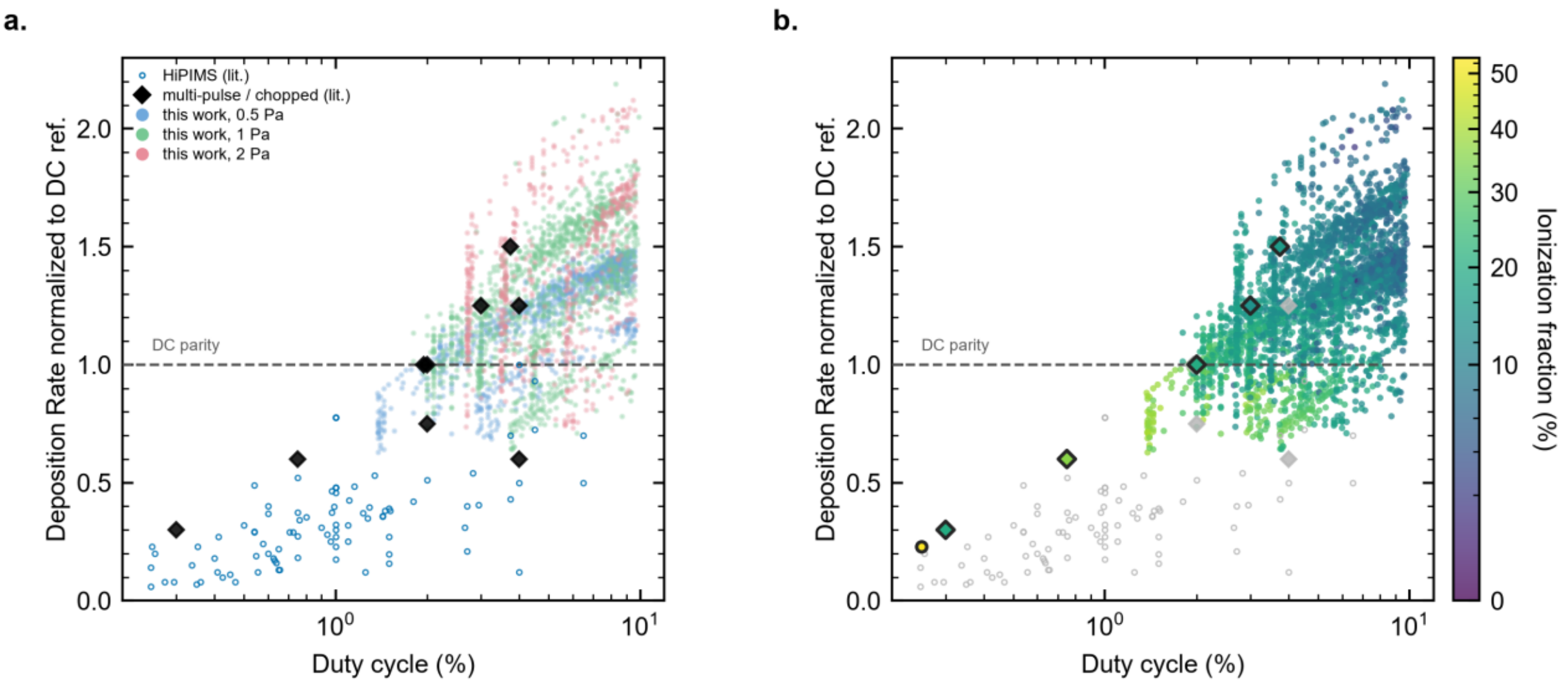

**Figure 3.** Overview of reported single- and multi-pulse HiPIMS processes with reference DC depositions rates. The HiPIMS processes are normalized to their DC dep-rates and plotted against their duty cycle. Single-pulse HiPIMS literature is shown in open blue circles and multi-pulse/chopped HiPIMS in black diamonds. The black dashed line indicates parity with DC deposition rates. In **a.** Points measured in this work colored by working pressure. **b.** The same conditions colored by ionized flux fraction on a square-root color scale. Literature points without a reported ionized flux fraction are greyed. Pulsed-DC reference depositions as well as additional campaigns run covering duty cycles of 10 – 50 % are shown in **Fig. S3**. Literature values span Cu, Ti, and Cr at working pressures of 0.07–2.7 Pa and several different magnet configurations.

**Figure 3** shows an overview of reported works that include a reference DC deposition rate, in both single-pulse HiPIMS and multi-pulse/chopped HiPIMS. This data is shown against their duty cycles, which highlights most acutely the different process regimes covered in HiPIMS. These deposition rates are all normalized to their DC references, with the black dashed line at 1 indicating 'break-even'. The measured points of this work are colored according to their working pressures in **Figure 3a** and according to their ionization flux fractions in **Figure 3b**. Reported values of single-pulse HiPIMS processes use duty cycles below about 5 % with most DC dep-rate ratios between 0.1 and 0.7,[3,6,20,29,40,41] mid-frequency pulsed-DC runs near DC parity at duty cycles of 50 % and above,[14,15,20] and multi-pulse reports only cross DC parity near a duty cycle of 2 %.[21,22,30] Our optimized processes reach deposition rates up to 1.5–2.2 times the DC rate at duty cycles of up to 10 %.

Our data follows a similar trend to the multi-pulse HiPIMS work published by *Hnilica et al.* in both the increasing deposition rate trends and the ionization flux fractions. A general trend towards increasing dep-rate over parity with DC is observed for increasing the duty cycle of the processes. Although there appears to be a point of diminishing returns followed by a decline at larger duty cycles, see **Figure S3** where we append additional campaigns run in a 10–50% duty cycle regime as well as references to pulsed-DC deposition rates.[15,20]

## 2.2 HiPIMS pulse packets at the pareto front

**Figure 4a–c** show, for each working pressure at 2.2 $Wcm^{-2}$, five validated-front processes with equally spaced pulse packets recorded for the accessible ionization range of those campaigns. In general, as packets move along the front they go towards higher duty cycles, trading packet length against peak-current density. At low ionization flux fractions the front becomes a long train of weak, flat pulses while at high ionization flux fractions, single, short pulses at three to five times the current-density (**Fig. S4 – S9**) are present.

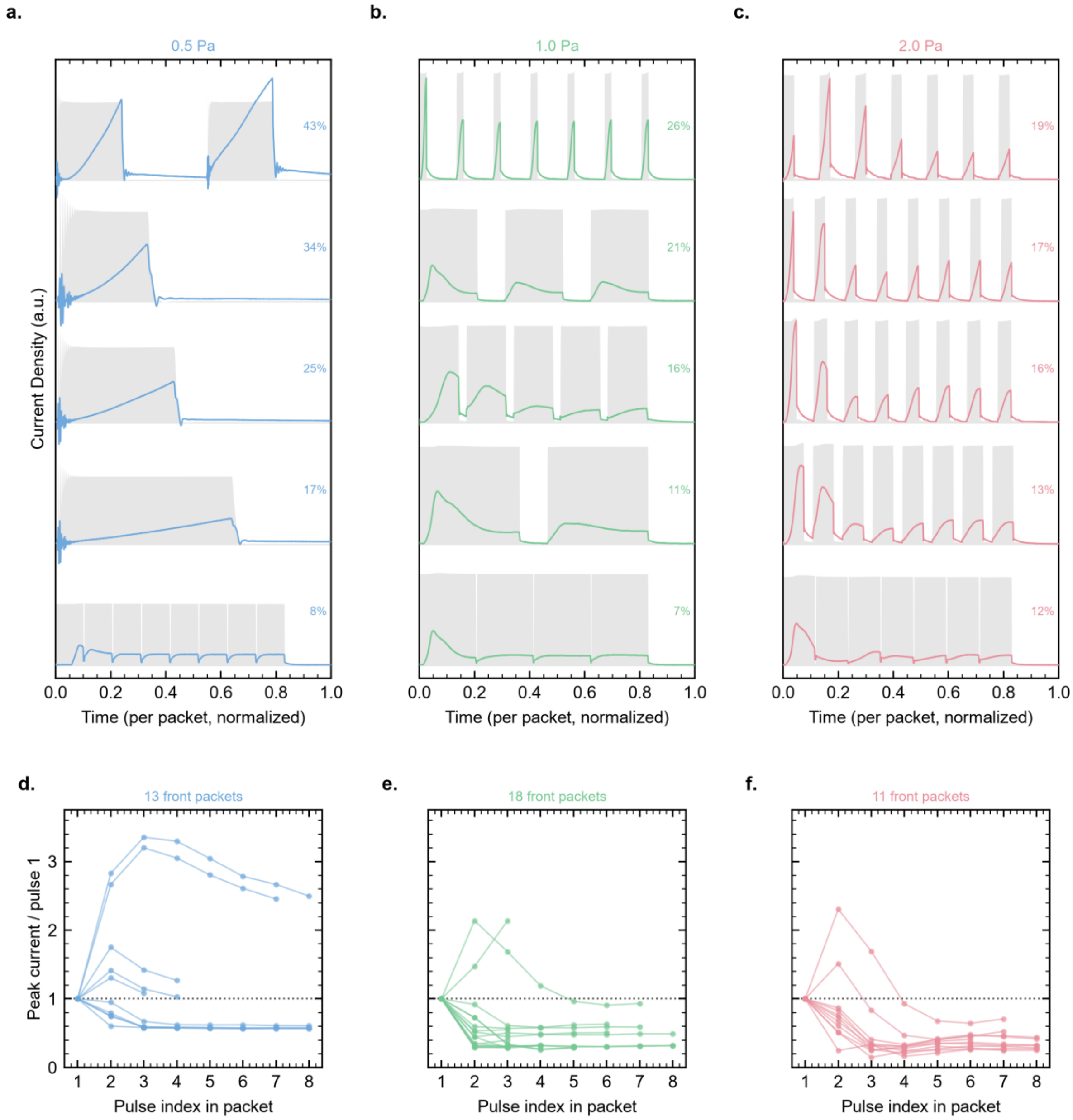

**Figure 4.** Front packet waveforms at 2.2 $Wcm^{-2}$. **a–c.** Five validated-front processes per pressure, chosen evenly across the accessible ionization range (low ionization at the bottom) with each trace normalized to its own packet duration. Shaded in grey are the voltage pulse on-times and to the right of each front packet is the measured ionization flux fraction. In **d–f.** The peak-current density of each pulse in the packet is normalized to the first pulse in that packet. The campaign run at 4.4 $Wcm^{-2}$ is shown in **Figure S10**.

The current density waveforms show a strong dependency on pressure. At 0.5 Pa the front is dominated by single pulses (**Figure 2c**) progressing more towards multi-pulse packets as the ionization flux fractions drop. At higher pressures, most points on the front are comprised of multi-pulse processes. Here, the first pulse ignites to high peak-current densities, rarefying the gas around it, followed by a decay into a stable plateau. These late pulses see thinner, partly pre-ionized gas and draw a significantly lower current at the same operating voltages. The 4.4 $Wcm^{-2}$ campaigns behave the same way (**Fig. S10**). This is seen more clearly for every front waveform, where the peak-current of every pulse in the packet is normalized to the peak-current of its first pulse, shown in **Figure 4d-f**.

Note that adding pulses to a packet is a proxy for extending pulse lengths and it is difficult to tell if the acquisition function during MOBO is optimizing for pulse packets or for longer single pulses. However, there is a strong physical suggestion that optimization is for single pulses. Typically, an increase in deposition rates is observed as the pulse width of a HiPIMS process decreases, due to the reduced effects of back-attraction. This comes from the ions generated in the plasma, which are attracted to the high negative voltage of the sputter target. However, this influence was seen to be negligible at pulse-widths below ~100 µs.[36] For front-facing pulse processes with <20% ionization flux fractions, we observe an average pulse-width of ≈170 µs (88–218 µs across all six campaigns), with total packet on-times of 660–1370 µs.

To further evaluate this, we ran 204 processes comparing single pulses to pulse packets with identical on times. The conglomerate of this data is shown in **Figure S11** with an example shown in **Figure 5a**, where a comparison of a single pulse to a pulse packet of 5 is shown. Both the deposition rates and ionization rates of this example pulse are nearly identical, within the error of the measurement itself. Only minor differences can be seen between the pulse packets and the single pulses, depending on the on-time of the packets (**Figure S11**).

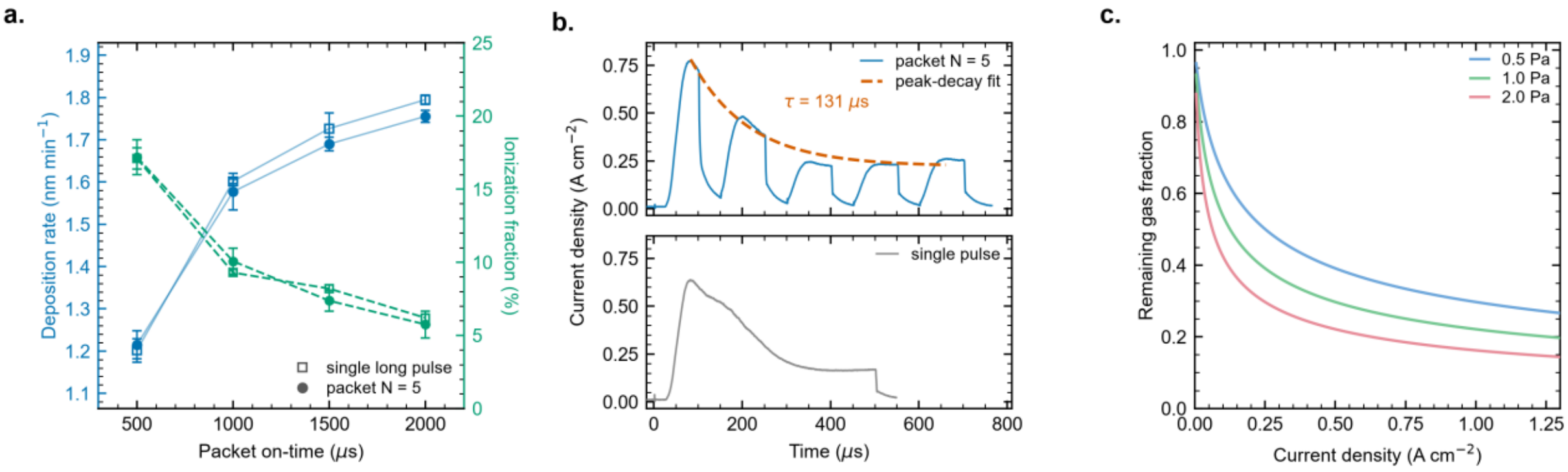

**Figure 5. a.** a pulse packet of five is compared to a single pulse of equivalent on-time, with negligible differences in ionized flux fraction (green) and deposition rates (blue). In **b.** a pulse packet is shown on the top panel in blue, with an exponential decay of the peak-current visible, fit with an orange dashed line. This is an effect of gas rarefaction, with gas continuing to be pushed out even between the pulses. An equivalent single pulse is shown below it in grey, with a similar decay, plateauing at similar point in time. In **c.** the conduction-limited rarefaction model from *Rossnagel*[25] was used to show the trend of rarefied gas as it depends on peak-current density. This was evaluated at each working pressure shown in this work using *Rossnagel's* reported constants and Ti sputter-yields from *Anders et al.*[5].

The reason for these similarities could be related to a physical transport limit of the rarefied gas. An initial, strong ignition pulse starts the rarefaction process, which continues to rarefy, regardless of subsequent pulses firing or not. This can be seen in **Figure 5b**, where a pulse packet is shown, experiencing the exponential decay of peak-current density. Here, a time constant of 131 µs can be fit, indicative of the rarefaction timescale. The relationship rarefaction has with peak-current in HiPIMS processes is well understood, and by adopting a conduction-limited rarefaction model, developed by *Rossnagel et al.*, the remaining gas fraction after the pulse against peak-current density is plotted in

**Figure 5c.** The first pulse is responsible for most of the gas rarefaction and every later pulse sputters into the channel it opens. At pulse packet gaps of 10 to 100 µs gas refill becomes negligible as it occurs at much larger timescale.[27,28]

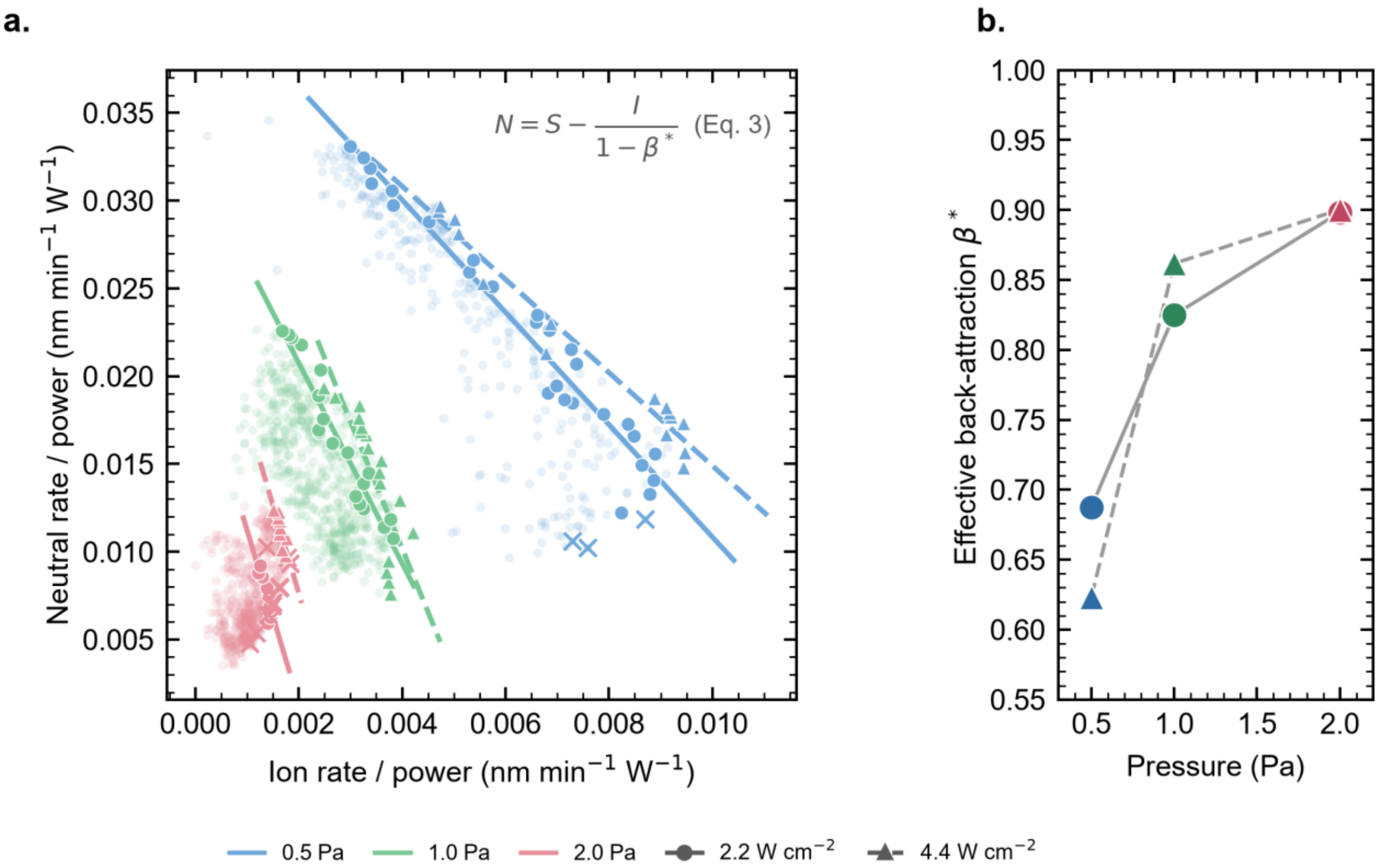

**Figure 6.** Neutral and ion deposition rates were normalized to power, shown in **a.** for 4.4 Wcm$^{-2}$ (dashed lines, triangles) and 2.2 Wcm$^{-2}$ (solid lines, circles). A linear fit of the Pareto front is used to estimate the effective back-attraction the front collapses to, based on the equations described in the main text, with β* = 1 − 1/k where k is the slope of the linear fit. Outliers beyond 2.5 standard deviations are excluded marked by an X. In **b.** the fitted β* is plotted against working pressure, showing an expected increase in effective back-attraction as working pressures rise.

### 2.3 The rate per watt is limited by process pressure

**Figure 6a** shows the deposition rates per watt for both neutrals and ions for the 2.2 and 4.4 Wcm$^{-2}$ campaigns. Remarkably the tradeoff between the normalized neutral and ion rates is almost linear, with different slopes for different pressure regimes. We attribute this to transport limits, set by gas rarefaction. A higher process gas density in front of the target would result in higher effective back-attraction and vice versa. This attribution comes from several assumptions. First, that the deposition rate of neutrals reduces by the number of ions produced in the discharge, such that:

$$N = S - f \qquad \text{Eq. 1}$$

Where N is the neutral flux, S is the total flux, and f is the ion flux. Next, of the ions that are generated, a certain amount are back-attracted to the target. The resulting ion-rate reaching the deposition meter can then be expressed as:

$$I = (1 - \beta^*)f \qquad \text{Eq. 2}$$

Where $I$ is the ion flux reaching the quartz crystal monitor and $\beta^*$ is the effective back-attraction. Substituting *Eq. 2* into *Eq. 1* we get:

$$N = S - I\frac{1}{(1-\beta^*)} \qquad \textit{Eq. 3}$$

The form of *Eq. 3* is linear, where the slope is $1/(1-\beta^*)$. In **Figure 6b**, back-attractions are fit using this linear equation and for 2.2 $Wcm^{-2}$ campaigns reaches a minimum effective back attraction of 0.90, 0.82, and 0.69 for working pressures of 2.0, 1.0, and 0.5 Pa respectively.

This decrease in effective back-attraction with pressure is an expected trend in HiPIMS, with both neutrals and ions experiencing reduced deposition rates as scatter events increase. In the case of ions, these trends are exacerbated, as scatter events that happen close to the target can severely increase the back-attraction probability of these ions.[41,42]

The collapse of the Pareto front to a linear regime strongly implies the transport limits described in this work, and the increase in effective back-attraction at higher working pressures is well-documented in literature. The ignition of a pulse packet expels a large part of the working gas in front of the target, and the ionization of subsequently sputtered flux then happens in a thinner medium. Ionization events then happen further away from the target, where the return field is weaker and an ion can escape more easily.[26,42,43] This rarefaction is highly dependent on the peak-current of the HiPIMS discharge, which drives the initial rarefaction towards an asymptote.

### 2.4 Modeling of the parameter space to guide future experiments

To elucidate the influence of individual process parameters on the Pareto front, we performed further sampling to allow for a more comprehensive coverage of the process parameter space up to the Pareto front. To this end we carried out an additional campaign at 1.0 Pa and 2.2 $Wcm^{-2}$. This was done using a standard Bayesian exploration algorithm described in *Experimental methods*, which evaluates the parameter space by steering each measurement toward the recipe the model is least certain about, driving the model's uncertainty down across the whole space. This falling uncertainty over sample number is shown in **Figure S12**. **Figure 7a** shows this campaign with its measured points in grey. A GPR was then fit to this space, and cross-fold validation was performed for both deposition rate and ionization flux fractions. A representative parity plot is shown in **Figure 7b**, showing a high accuracy with $R^2$ of 0.99 and 0.85 for the deposition rate and ionization respectively. The measured ionization rates were expected to have lower accuracy then the deposition rates, following the spread of drift measurements reported in **Figure S13**. Using the GPR we can model Pareto fronts of different pulse numbers, shown in different colors in **Figure 7a**. The importance of pulse count increases as the ionized flux fractions increase. However, we would like to emphasize that adding pulses to the packet is also a proxy for increasing pulse-length and the convolution of pulse-packet number and pulse on-time cannot be resolved by this visual.

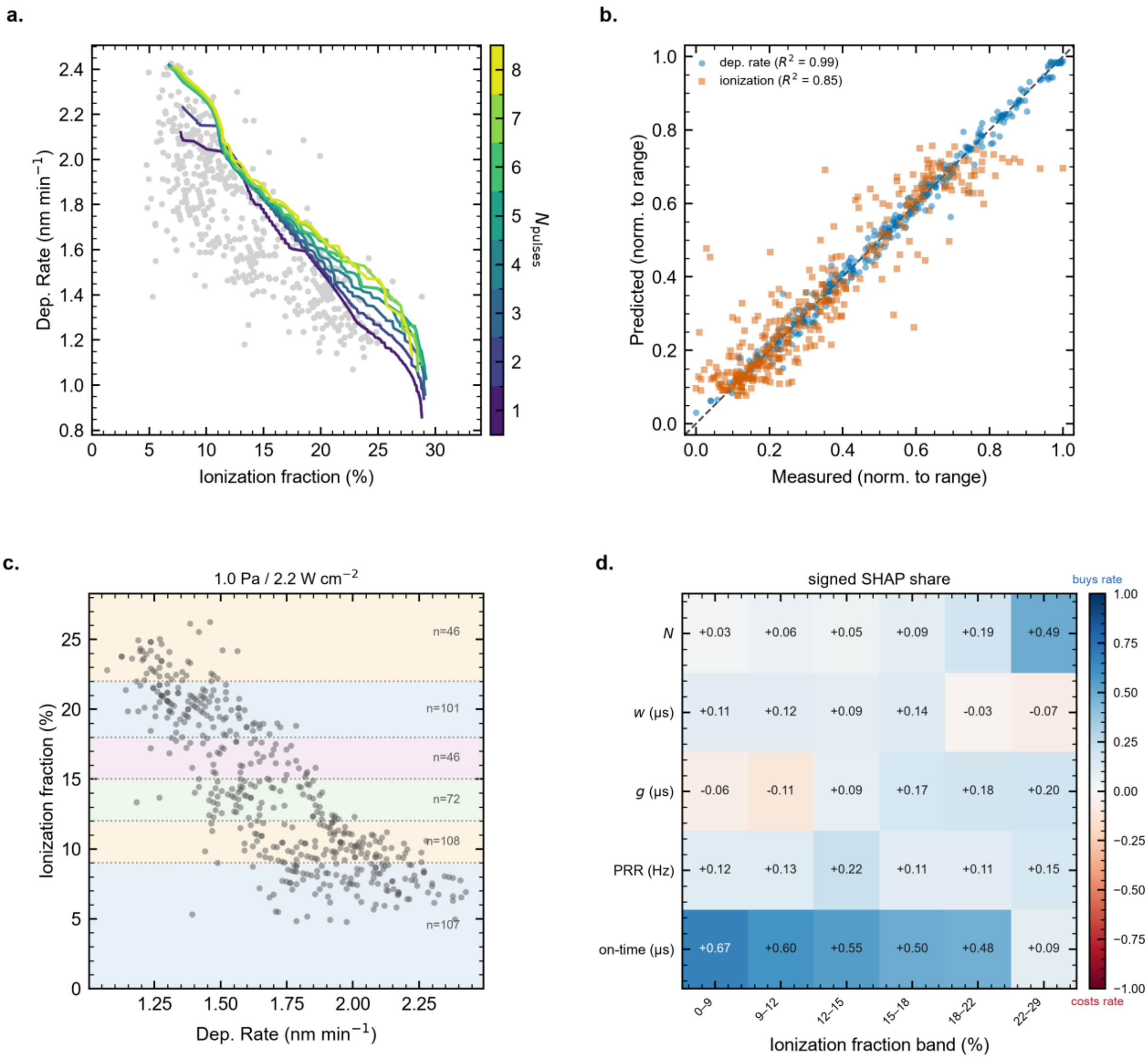

**Figure 7.** Using Bayesian optimization algorithms, the uncertainty across the parameter space of the 1.0 Pa, 2.2 Wcm$^{-2}$ campaign was minimized. Subsequently, a GPR trained on the dataset was used to model Pareto fronts at different pulse counts, shown in **a.** in different colors. In **b.** parity plot of cross-fold validation of the GPR for both deposition rate and ionization flux fractions is shown. In **c.** the campaign is split into ionization bins, with the number of processes in each bin labeled in grey on the right. These bins are used to normalize SHAP values shown in panel **d.** These normalized SHAP values indicate if an increase in those parameters raise the deposition rate (blue) or reduces it (red).

To resolve these effects, we split the campaign into different ionized flux fraction bins, shown in **Figure 7c**. This allows us to evaluate the feature importance of all varied process parameters in different ionized flux fraction regimes. Based on the GPR we calculate SHAP values for each ionized flux fraction regime, described in the *Experimental Methods*. **Figure 7d** shows these normalized SHAP values for the different process parameters across each ionized flux fraction regime. SHAP is a statistical analysis method which can help to deconvolute the effects of process parameters. The increase in pulse-count over on-times at higher ionized flux fractions indicate that the observed spreading in the GPR at ionizations <22% is dominantly from increasing on-times, related to the proxy previously mentioned.

This SHAP value matrix can be regarded as a guide for experimentalists to choose the most promising parameters to optimize deposition rate at a given ionized flux fraction. Positive numbers, shaded in blue, indicate that an increase in that parameter positively affects the deposition rate, and vice-versa (negative numbers shaded in red). The total values are indicative of the percentage that each parameter is expected to affect the deposition rate. As an example, the SHAP value for on-time of 0.67 in the ionized flux fraction regime of 0-9 % means that 67 % of the deposition rate increase is attributed to the on-time. The sum of the absolute value of each column equals 1 as a result. Across all considered ionized flux fractions regimes we find a remarkable trend: only pulse count and on-time show significant correlation with deposition rate. On-time is most important at lower ion fractions, whereas pulse count only begins to dominate at ion fractions >22%.

Practically, the results described in this work imply an important design principle for sputter processes. For moderate ionization flux fractions (between 5–20% for a working pressure of 1.0 Pa), deposition rates can be maximized by moving towards ms-scale single pulses. Important considerations include the peak-current density of that pulse, which should be high enough to maximally rarefy the gas channel. However, most of the energy of the pulse should go into its trailing end, which plateaus at a lower current value and pushes ions and neutrals through the rarefied channel. Optimization of on-time and frequency is key to achieve this. For higher ionization flux fractions, pulse-packeting remains a viable option where expected benefits would be seen for pulses shorter than ~100 µs. Likely, even at these higher ion counts, an interplay between gas rarefaction and back-attraction is at work. Note that the accessible ion counts vary strongly with pressure, but general trends are observed for all three working pressures.

## 3. Conclusion

High deposition rates are a key feature of sputter processes that have led to their widespread adoption. Historically, DC deposition rates have been viewed as optimal for high deposition rates in the industry. In this work, we are able to show deposition rates nearly twice as high as their DC references, which we accomplished using multi-objective Bayesian optimization (MOBO). This allowed us to find the Pareto front for the inherent tradeoff between high ion counts and high deposition rates. Surprisingly, for moderate ionization flux fractions, pulse packets do not seem to be effective, instead relying on ms-scale pulses and significantly higher duty cycles to reach the front. We hypothesize this is related to the maximal rarefaction of the gas channel between the sputter target and the substrate, set by the initial peak-current density, which sets a minimum effective back-attraction for the process.

This is a significant shift from current pulse design principles, operating in a little explored process space between HiPIMS and pulsed-DC sputtering. Importantly, while several applications rely on these high ion counts, there is significant interest in softer deposition processes. This is exemplified by the large body of work studying piezo- and ferro-electric materials, diffusion barriers, transparent conducting oxides, and conformal coatings with pulsed sputter processes.[7,12,44,45] The presented results are a significant step forward for the sputter deposition applications, using duty cycle and frequency in pulsed sputtering and giving physically motivated design principles to achieve higher than DC dep rates while maintaining ion counts.

## 4. Experimental methods

### 4.1 Deposition system

All experiments were performed at Empa in a custom-built sputter system, equipped with custom python control software. Sputtering from AJA A330 unbalanced magnetrons from with a single 3" diameter Ti target, provided by Kurt J. Lesker at 99.99% purity. On axis geometry, working distance 12 cm. Chamber pressure (0.5, 1.0, and 2.0 Pa) was set by changing the pumping speed, throttling the gate valve at a fixed argon flow of 50 sccm. An Ionautics® HiPSTER 6 power supply was used, equipped with a SiC transistor to allow for 100 kHz pulsing speeds.

An ion meter was fabricated by modifying a quartz crystal monitor (QCM), equipping it with a pair of neodymium magnets at the front to repel electrons and a low-pass filter to allow for a DC bias to be applied to the front surface of the QCM. This was built following guidelines from *Kubart et al.*[46] Biasing the front at +30 V repels ions, measuring only neutrals whilst grounding the front allows both neutrals and ions to deposit. In this way, ionization flux fractions were measured.

### 4.2 Active-learning campaign design

Pulse packet processes investigated in this work were designed symmetrically, meaning the spacing and width of each pulse in the packet is identical. Variable pulsing schemes were investigated but negligible differences were observed in the Pareto Front, this is shown in **Figure S14**. Each Pareto campaign was run using a dual-objective Pareto objective, maximizing deposition rate and ionized flux fraction simultaneously. The recommender is a two-phase meta-recommender with the first ten trials sampled randomly, followed by a BoTorch-based Bayesian recommender. Examples of the machine-learning algorithms used in this work are provided in *Supporting Information*. The surrogate is a composite of independent Gaussian-process models, one per objective, and its acquisition function is the q-log noisy expected hypervolume improvement with reference point (0, 0), optimized with ten restarts from 64 raw samples over the discrete search space of **Table S1**. A fixed drift reference was used, measuring an identical process every 20 processes to track evolution of the measurement equipment or target state, shown in **Figure S13**. Processes on the Pareto front were then validated, measuring them three times at the end of a campaign and ensuring that the standard deviation of their ionization and deposition rats fell below 2.5%. Most campaigns had acceptable drift, however two of the 4.4 $Wcm^{-2}$ experienced significant drift in the process. This is documented in **Figure S15** and led to the removal of many points in those campaigns and the reporting of several points on the front which were not validated.

### 4.3 Surrogate model and SHAP statistics

The per-pulse-count fronts of **Figure 7** come from a GPR trained per objective on the four process parameters (pulse count, spacing, width, and frequency). These are trained using scikit-learn Python packages and Matérn kernel with $\nu = 5/2$, an independent length scale per knob, and a white-noise term, with targets normalized. Model quality is reported as a five-fold cross-validation, with folds grouped by condition so that repeats of the same condition never straddle a train/test split.

SHAP values shown in **Figure 7** are calculated from gradient-boosted models (XGBoost, 500 trees, depth 4) of the deposition rate, trained separately inside each ionized-flux-fraction band. The signed SHAP values are then calculated by taking the mean SHAP values of the parameter and dividing by the sum of all SHAP values across all parameters. As such, the SHAP values represent the percent that the parameter influences the output. The direction, negative or positive, of the SHAP attribution is then determined by Spearman correlation coefficients of calculated SHAP values.

## Acknowledgements

This work was partially supported by different projects. N. Rodkey and S. Siol acknowledge funding from the Swiss National Science Foundation (SNSF) under grants 227945 and 10004403. F. Thelen and A. Ludwig acknowledge funding from the European Union through the European Research

Council (ERC) Synergy Grant project 101118768 (DEMI). Views and opinions expressed are however those of the authors only and do not necessarily reflect those of the European Union or ERCEA. Neither the European Union nor the granting authority can be held responsible for them. The authors would like to further acknowledge Manuel Kober-Czerny for helpful discussions and Sasa Vranjkovic for support with the design of components.

Supporting Information:

# Multi-objective Bayesian optimization reveals HiPIMS deposition rates exceeding direct current sputtering

Nathan Rodkey[1]*, Felix Thelen[2], Alfred Ludwig[2,3], Sebastian Siol[1]*

[1] Empa, Swiss Federal Laboratories for Materials Science and Technology, Laboratory 208, Überlandstrasse 129, 8600 Dübendorf, Switzerland

[2] Materials Discovery and Interfaces, Institute for Materials, Ruhr University Bochum, 44801 Bochum, Germany

[3] Center for Interface-Dominated High-Performance Materials (ZGH), Research Center Future Energy Materials and Systems (RC FEMS), Ruhr University Bochum, 44801 Bochum, Germany

*Corresponding authors: nathan.rodkey@empa.ch, sebastian.siol@empa.ch

## Conduction-limited rarefaction model

The gas-density decline with peak current density shown in **Figure 4c** of the main text is the conduction-limited rarefaction model of *Rossnagel*, evaluated with the constants ($E_a$ = 8 eV, σ = 6×10$^{-16}$ cm$^2$, f = 3, K the thermal conductivity of argon at 300 K) and the Ti sputter-yield fit of Anders (Y = 0.60 at 400 V):

$$n_H = \frac{-T_0 + \sqrt{T_0^2 + \beta I}}{\gamma I} \quad \textit{Eq. S1}$$

$$\beta = \frac{E_a Y \sigma n_0 T_0}{\pi f K} \quad \textit{Eq. S2}$$

$$\gamma = \frac{E_a Y \sigma}{2\pi f K} \quad \textit{Eq. S3}$$

which converges at high current to the limiting law:

$$n_H \rightarrow (\frac{2 n_0 T_0 \pi f K}{E_a Y \sigma})^{1/2} I^{-1/2} \quad \textit{Eq. S4}$$

Where $n_H$ and $n_0$ are the heated-zone and ambient gas densities, $T_0$ the ambient temperature, I the discharge current, Y the sputter yield of Ti, $E_a$ the average energy per sputtered atom, σ the momentum-transfer cross-section, f the extent of the heated zone in mean free paths, and K the thermal conductivity of argon.

**Table S1.** Lower and upper bounds used for pulse packets in the six campaigns shown in this work.

| Parameter | Lower | Upper | Step |
|---|---|---|---|
| Pulse count N | 1 | 8 | 1 |
| Pulse width w (µs) | 10 | 500 | 10 |
| Gap g (µs) | 10 | 100 | 10 |
| Repetition frequency PRR (Hz) | 50 | 2000 | 10 |

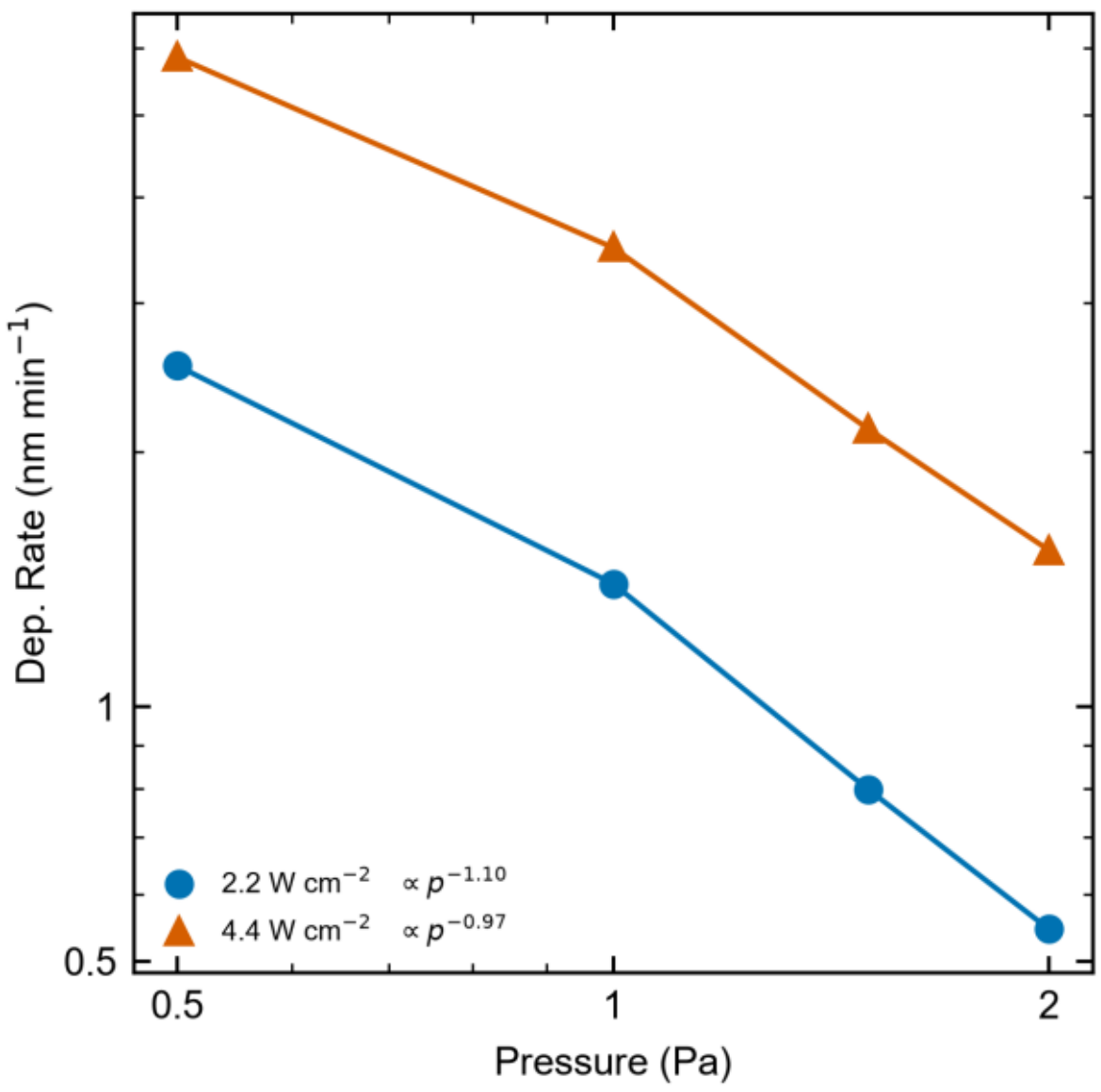

**Figure S1.** The DC reference against pressure. The measured DC deposition rate at both power densities on log axes, with the log-log neighbor interpolation every DC comparison in the main text uses; a single power law leaves 10 % residuals and reads low near 1 Pa.

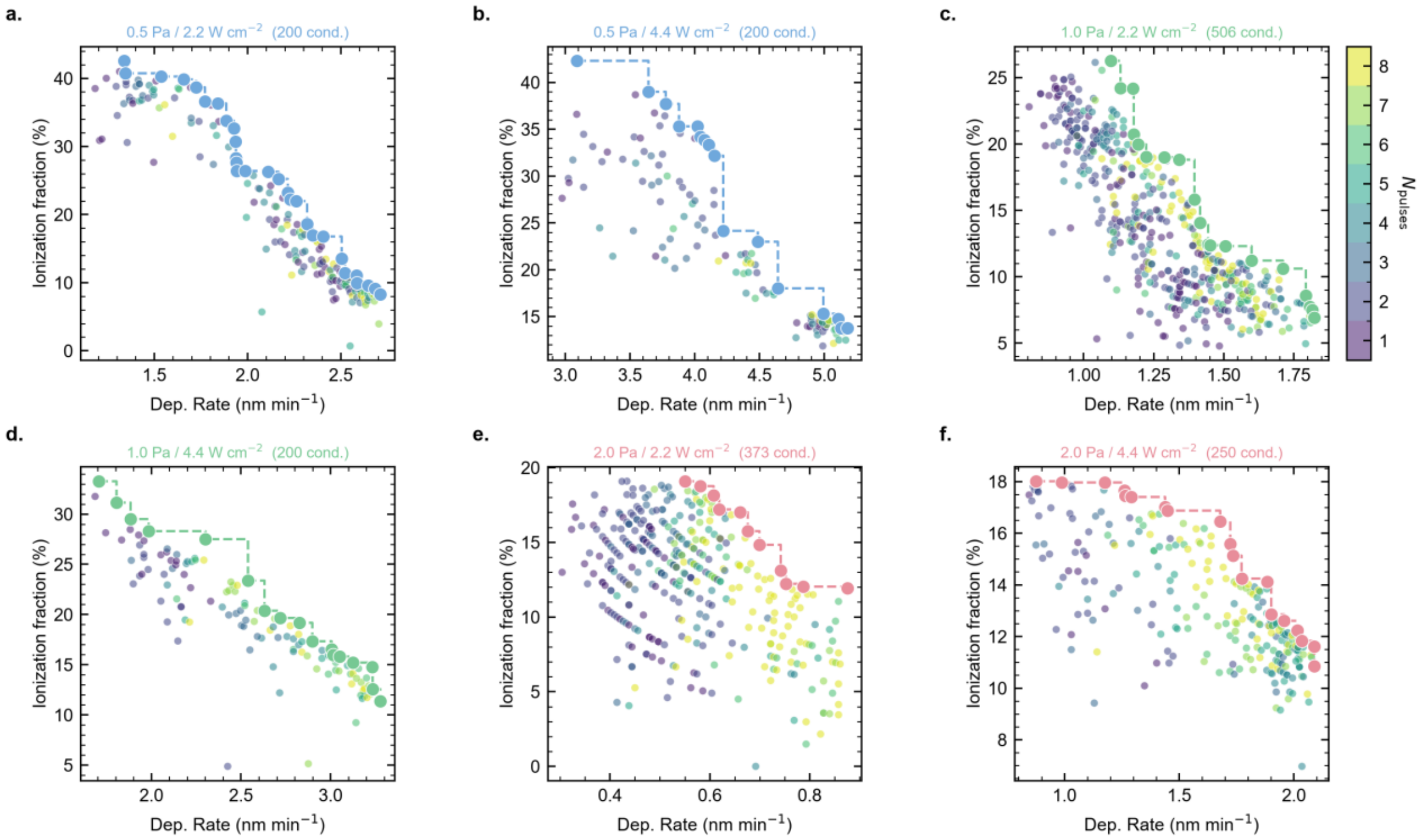

**Figure S2.** The six Pareto front campaigns studied in this work are shown above, color-coded by working pressure with 0.5 Pa in blue, 1.0 Pa in green, and 2.0 Pa in red. The validated Pareto front, as described in the *Experimental Methods*, is shown by the solid-colored dots, connected by a dashed line. Points are color-coded by pulse count, where a clear separation in this pulse count is observed at higher working pressures.

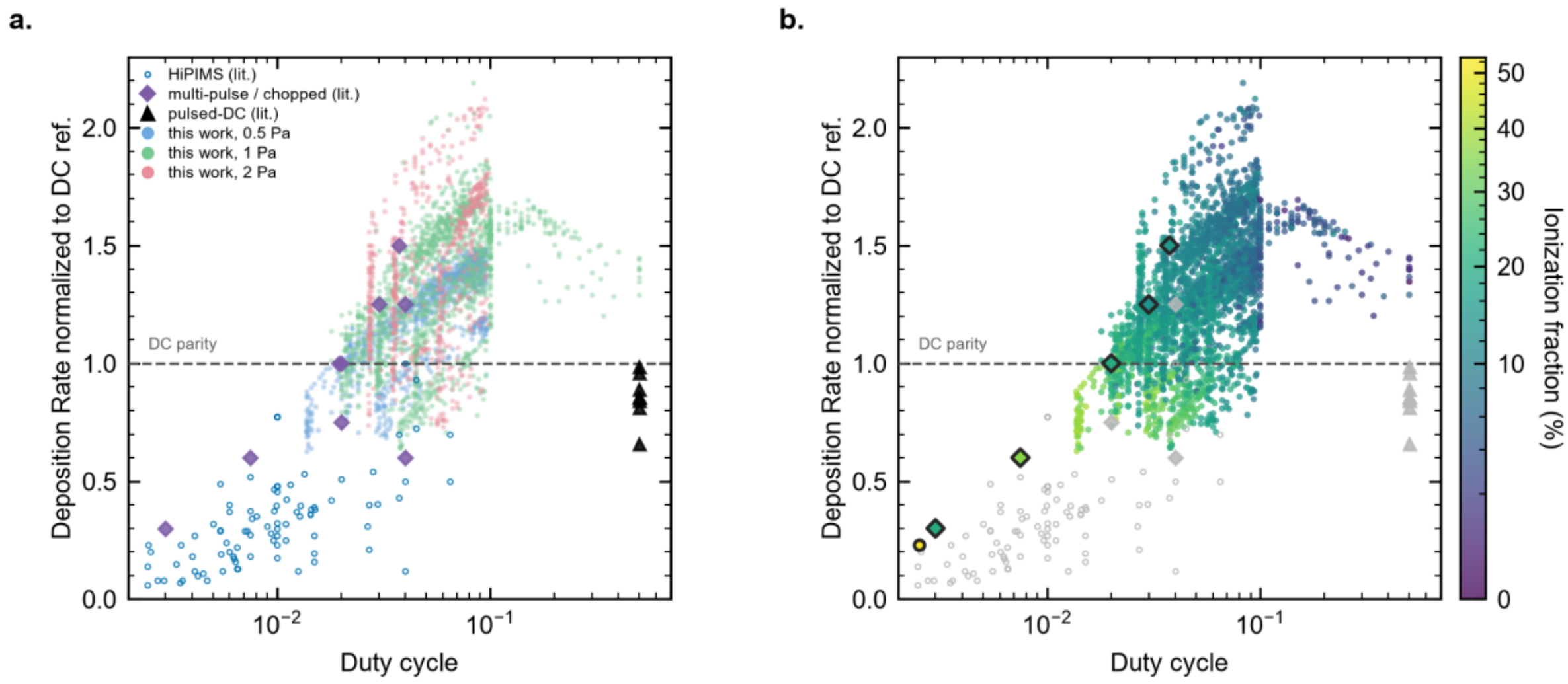

**Figure S3.** Single-pulse, multi-pulse HiPIMS, and pulsed-DC processes that have reference DC depositions rates are shown above. The HiPIMS processes are normalized to their DC dep-rates and plotted against their duty cycle. Single-pulse HiPIMS literature is shown in open blue circles and multi-pulse/chopped HiPIMS in purple diamonds. The black dashed line indicates parity with DC deposition rates. In **a.** Points measured in this work colored by working pressure. **b.** The same conditions colored by ionized flux fraction on a square-root color scale. Literature points without a reported ionized flux fraction are greyed. Pulsed-DC reference depositions as well as additional campaigns run covering duty cycles of 10–50 % are shown in **Fig. S3**. An additional campaign done at higher duty cycles (10–50%), at 1.0 Pa is shown.

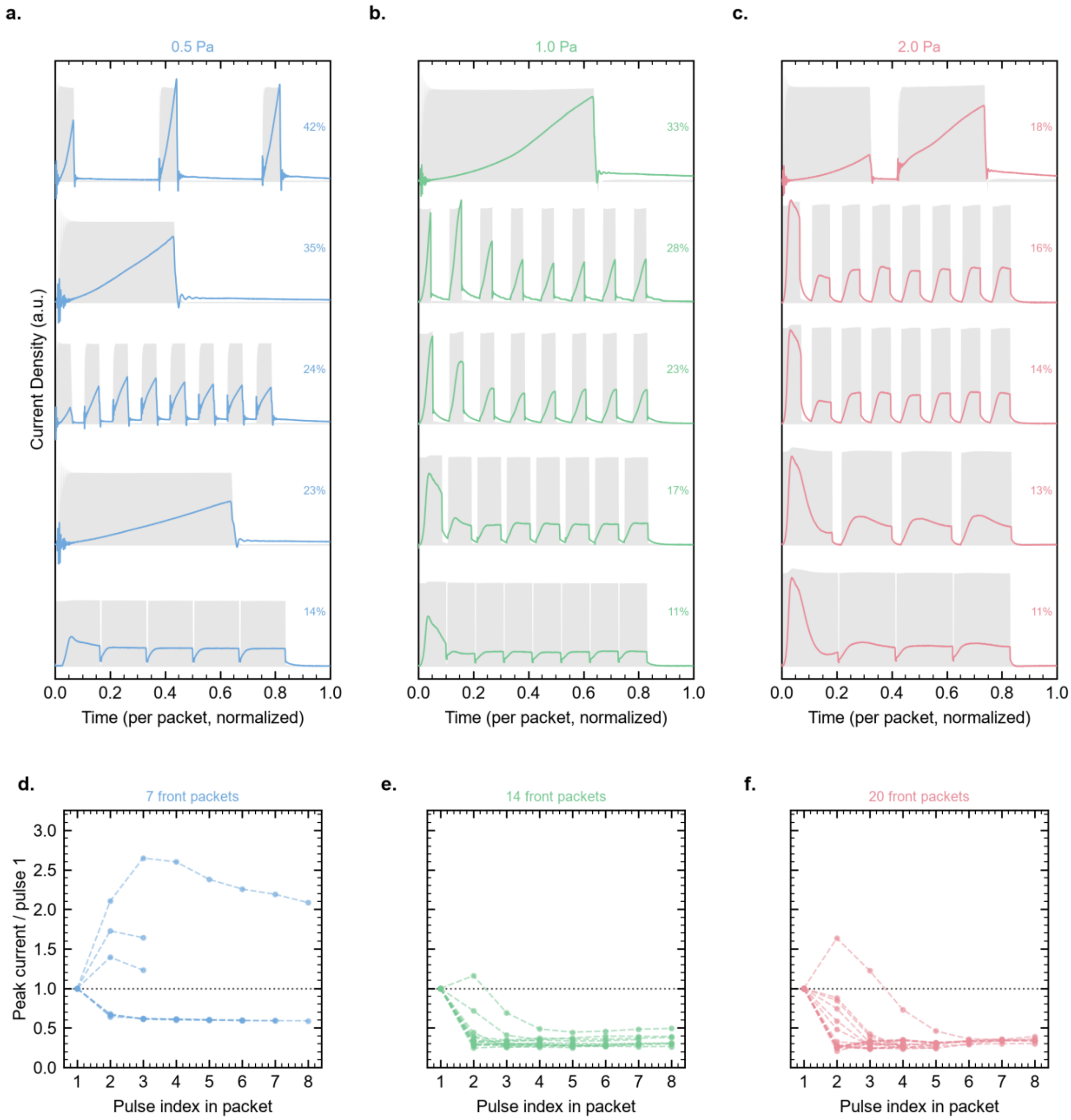

**Figure S4.** Front packet waveforms at 4.4 $Wcm^{-2}$. **a–c.** Five validated-front packets per pressure, chosen evenly across the accessible ionization range (low ionization at the bottom) with each trace normalized to its own packet duration. Shaded in grey are the voltage pulse on-times and to the right of each front packet is the measured ionization flux fraction. In **d–f.** The peak-current density of each pulse in the packet is normalized to the first pulse in that packet.

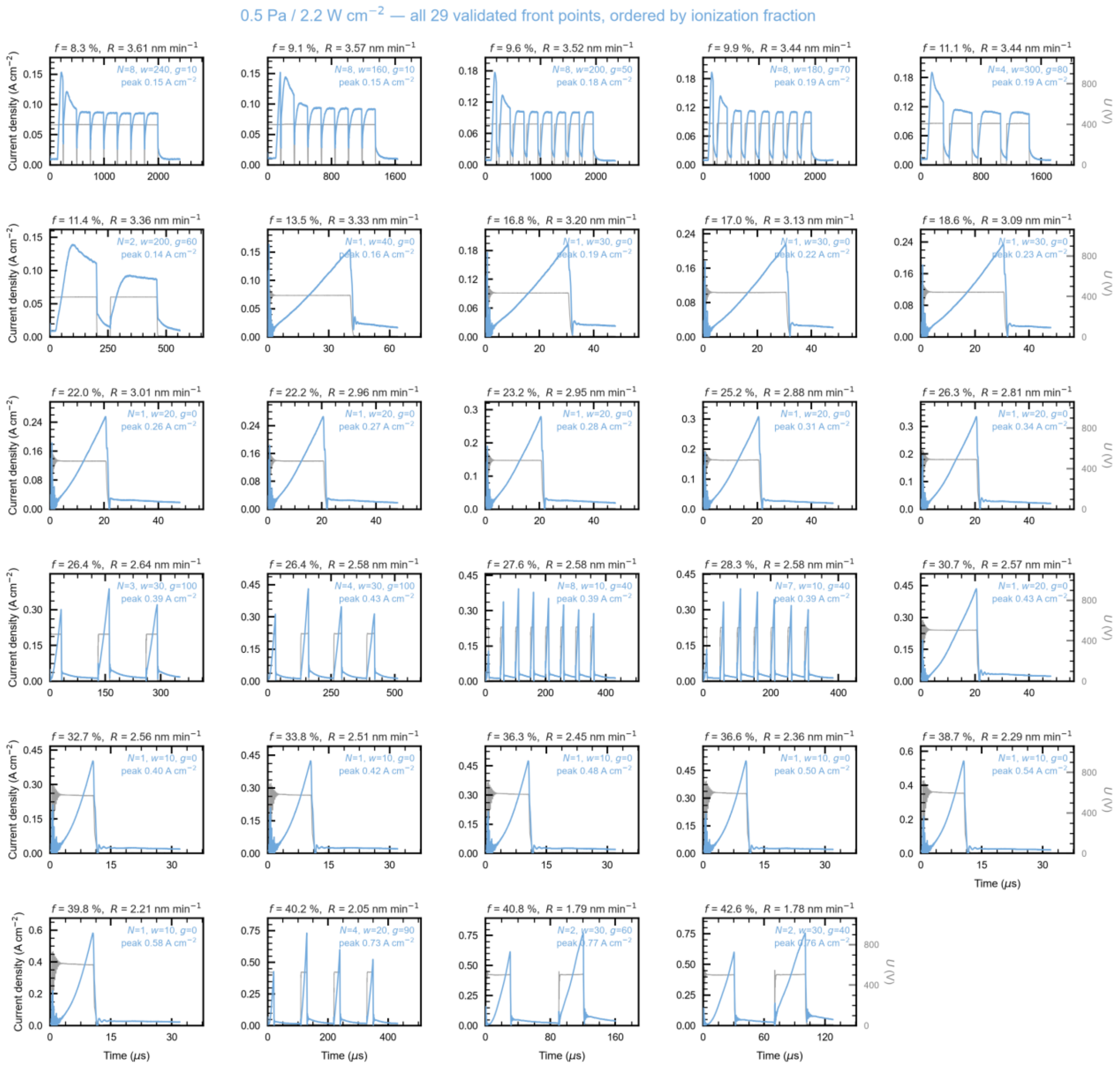

**Figure S5.** Every validated front point of the 0.5 Pa, 2.2 Wcm⁻² campaign, ordered by ionized flux fraction starting with the lowest ionization flux fractions at the top left. These ionization flux fractions are labeled above each graph, alongside their deposition rates. The pulse count, gap between pulses, pulse-widths, and peak-current densities are labeled within each panel. The peak-current discharge is shown in blue with its corresponding voltage discharge shown in grey.

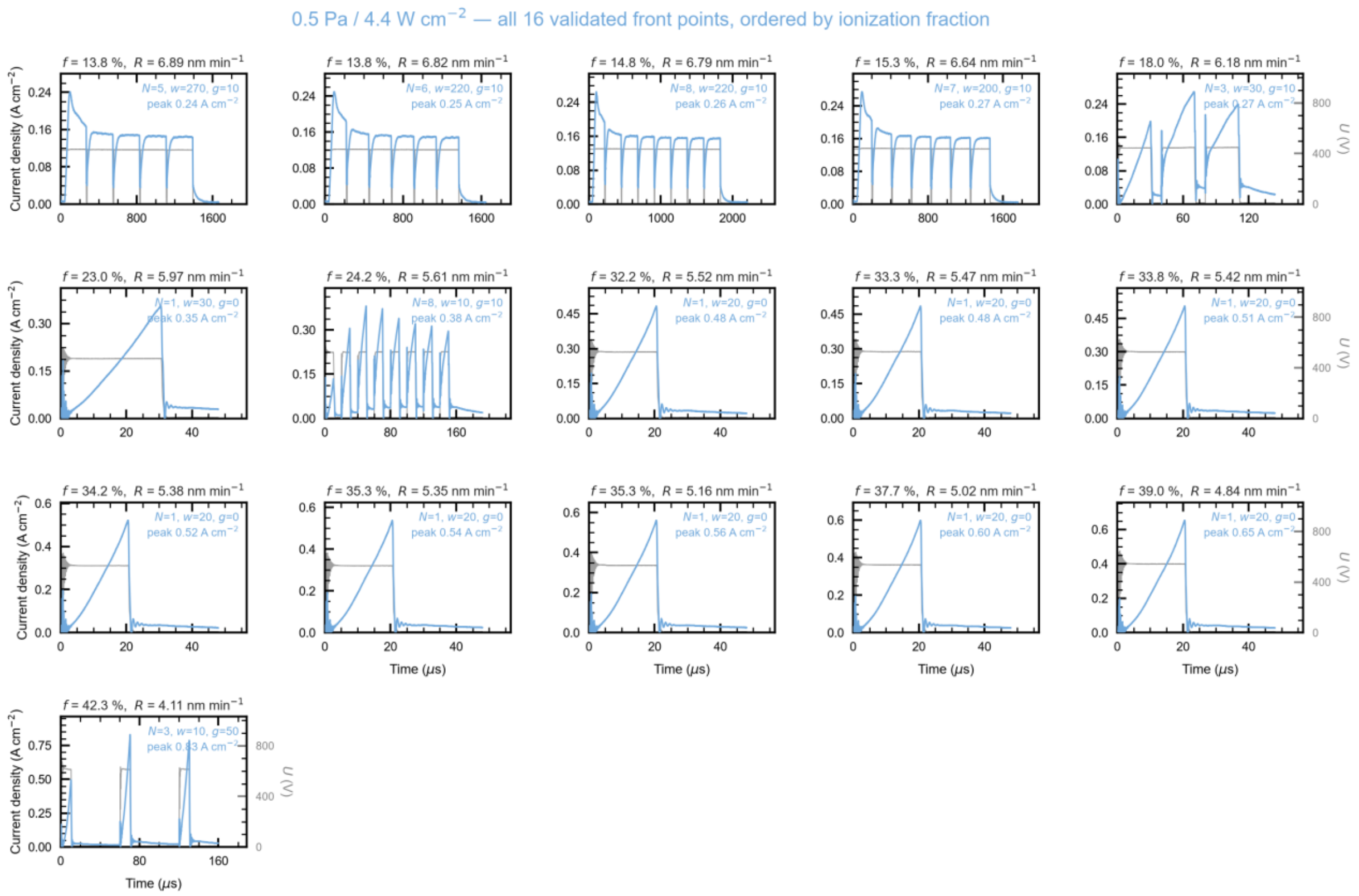

**Figure S6.** Every validated front point of the 0.5 Pa, 4.4 Wcm⁻² campaign, ordered by ionized flux fraction starting with the lowest ionization flux fractions at the top left. These ionization flux fractions are labeled above each graph, alongside their deposition rates. The pulse count, gap between pulses, pulse-widths, and peak-current densities are labeled within each panel. The peak-current discharge is shown in blue with its corresponding voltage discharge shown in grey.

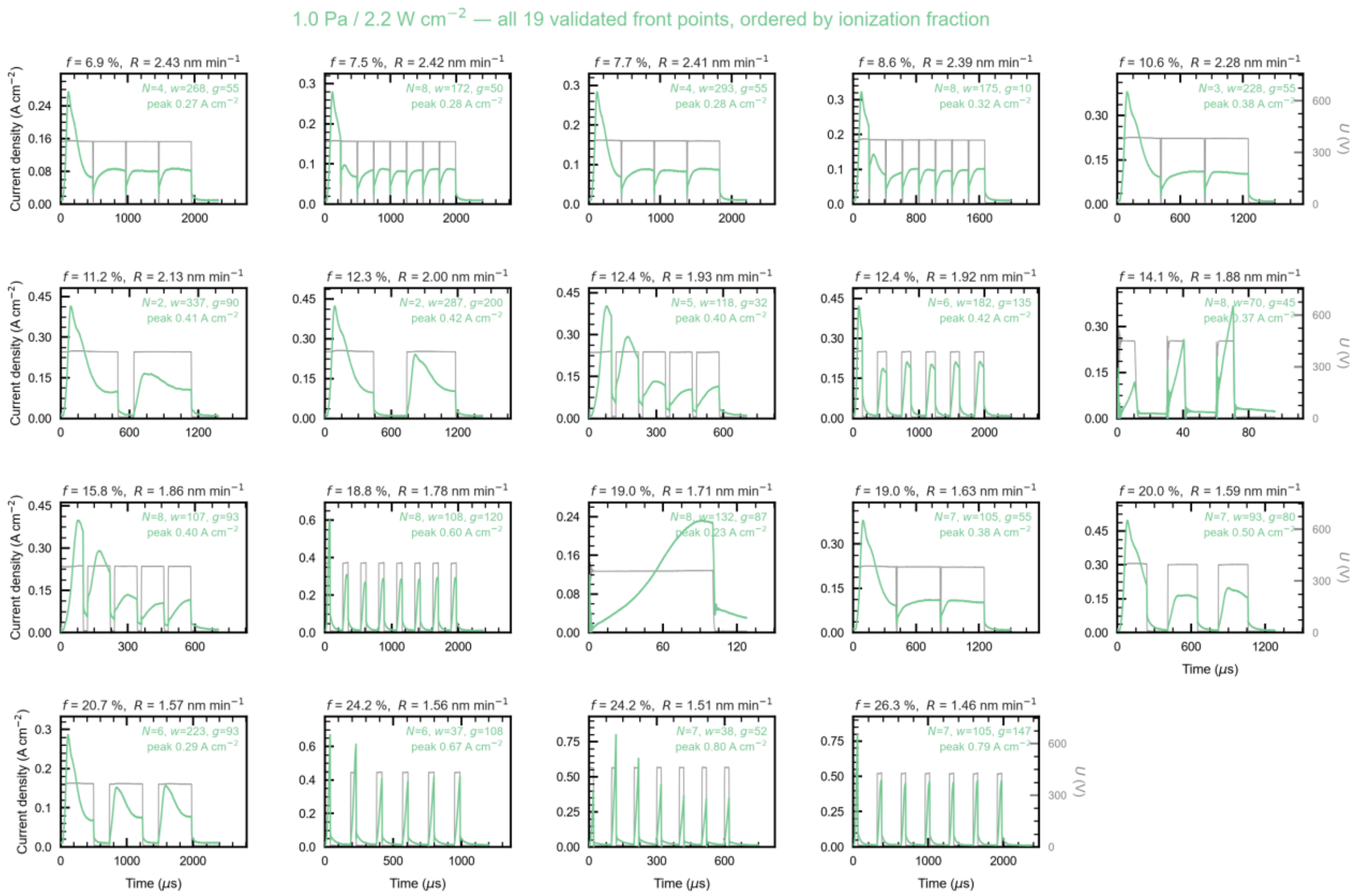

**Figure S7.** Every validated front point of the 1.0 Pa, 2.2 Wcm⁻² campaign, ordered by ionized flux fraction starting with the lowest ionization flux fractions at the top left. These ionization flux fractions are labeled above each graph, alongside their deposition rates. The pulse count, gap between pulses, pulse-widths, and peak-current densities are labeled within each panel. The peak-current discharge is shown in blue with its corresponding voltage discharge shown in grey.

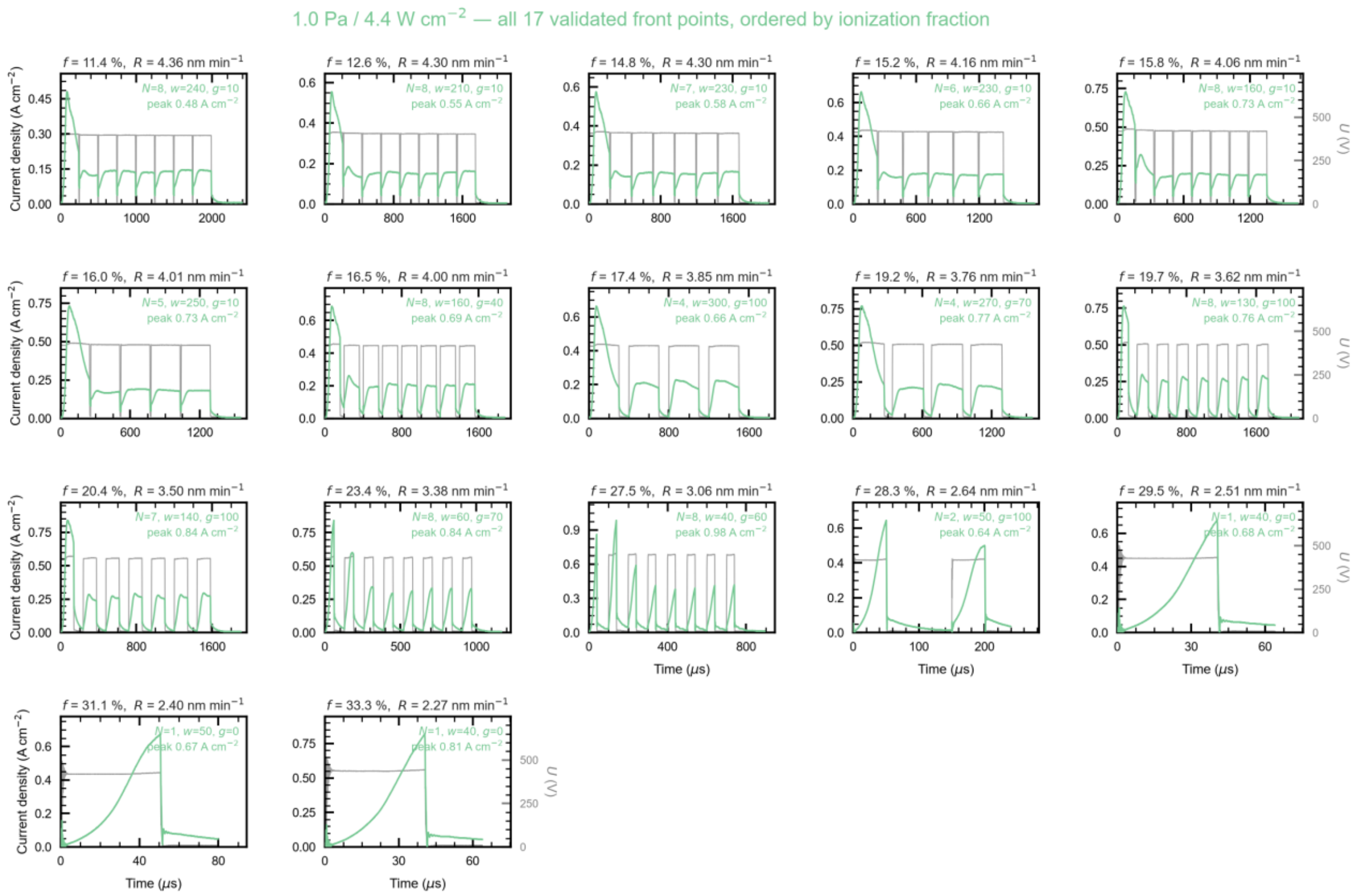

**Figure S8.** Every validated front point of the 1.0 Pa, 4.4 Wcm⁻² campaign, ordered by ionized flux fraction starting with the lowest ionization flux fractions at the top left. These ionization flux fractions are labeled above each graph, alongside their deposition rates. The pulse count, gap between pulses, pulse-widths, and peak-current densities are labeled within each panel. The peak-current discharge is shown in blue with its corresponding voltage discharge shown in grey.

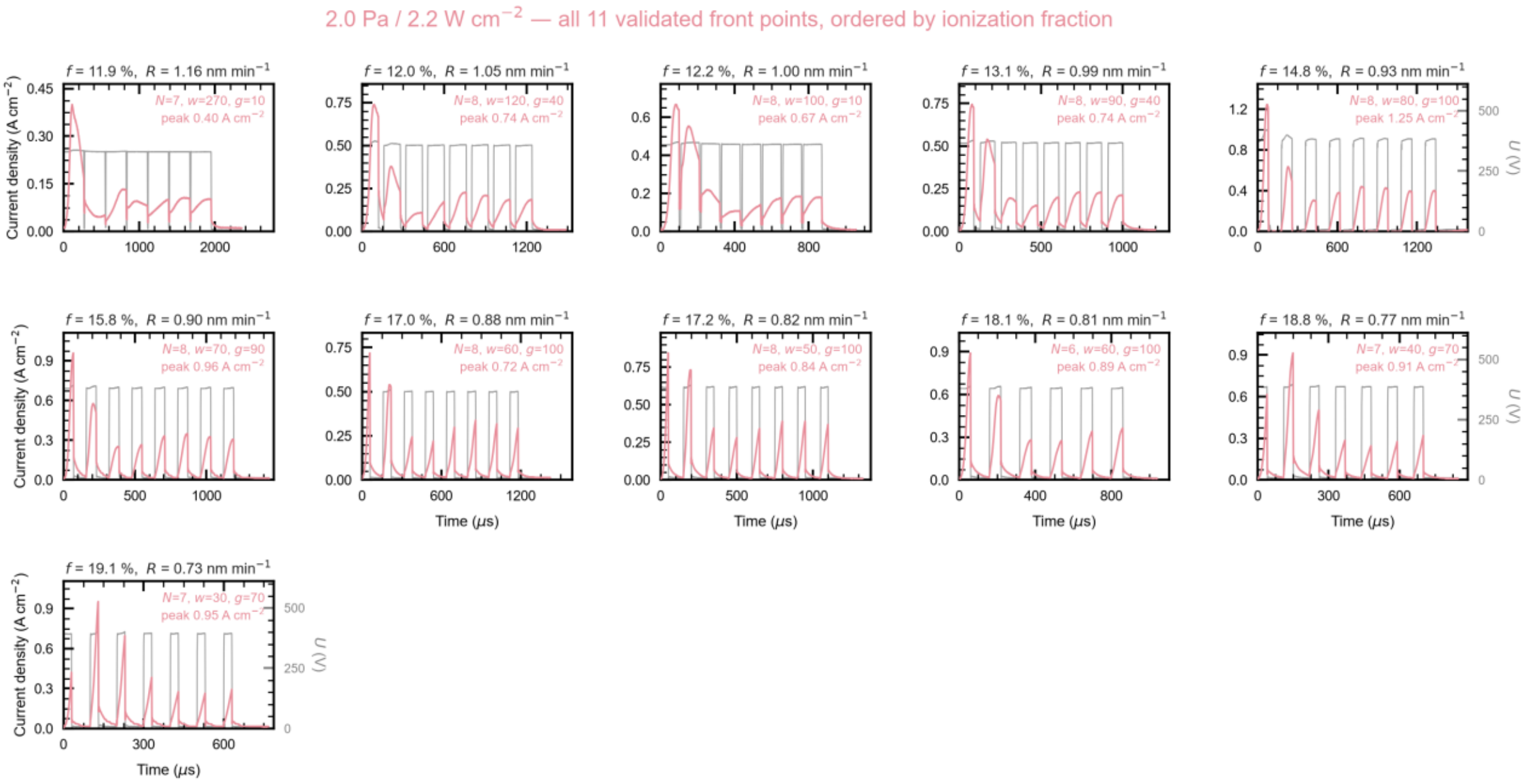

**Figure S9.** Every validated front point of the 2.0 Pa, 2.2 Wcm⁻² campaign, ordered by ionized flux fraction starting with the lowest ionization flux fractions at the top left. These ionization flux fractions are labeled above each graph, alongside their deposition rates. The pulse count, gap between pulses, pulse-widths, and peak-current densities are labeled within each panel. The peak-current discharge is shown in blue with its corresponding voltage discharge shown in grey.

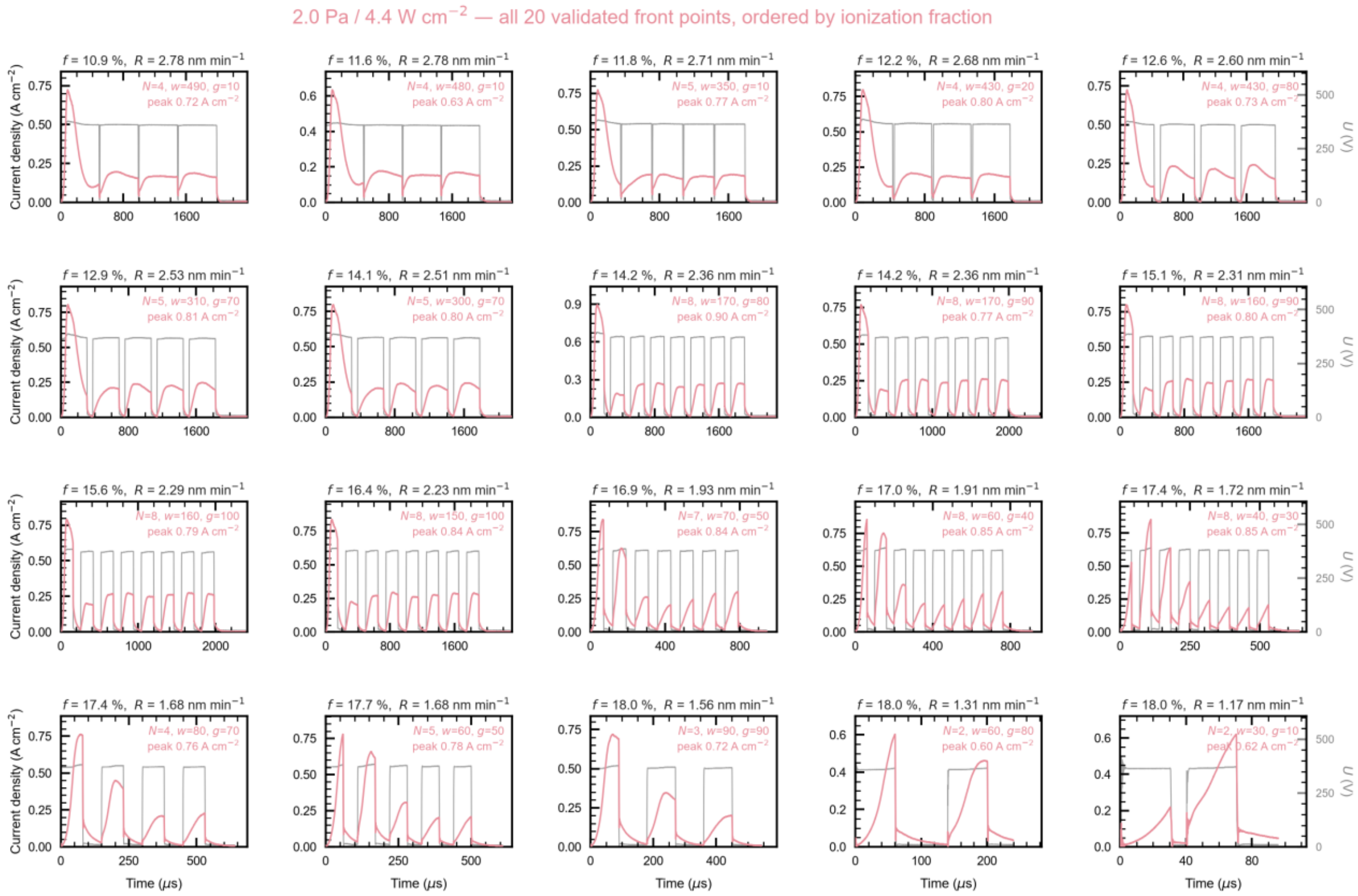

**Figure S10.** Every validated front point of the 2.0 Pa, 4.4 Wcm⁻² campaign, ordered by ionized flux fraction starting with the lowest ionization flux fractions at the top left. These ionization flux fractions are labeled above each graph, alongside their deposition rates. The pulse count, gap between pulses, pulse-widths, and peak-current densities are labeled within each panel. The peak-current discharge is shown in blue with its corresponding voltage discharge shown in grey.

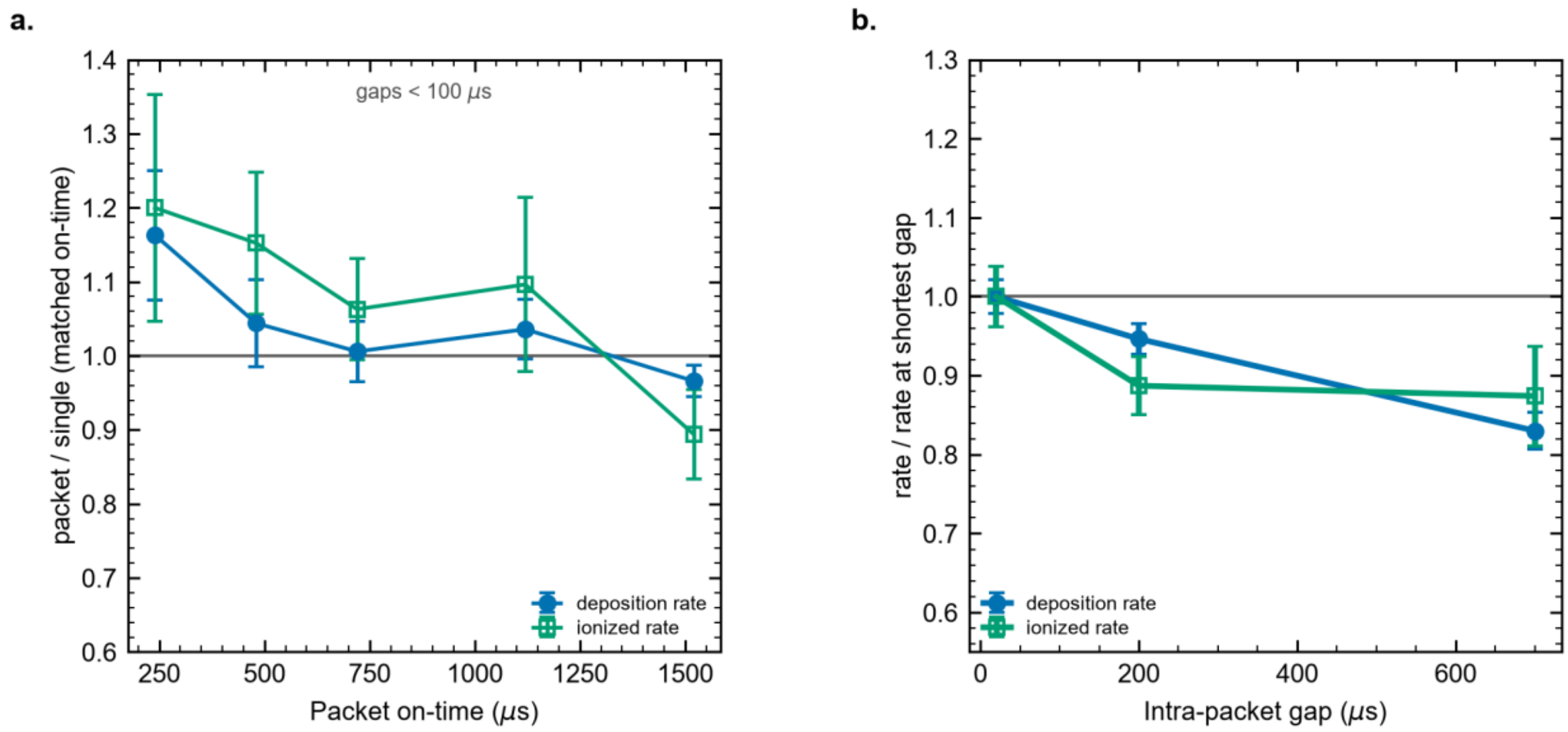

**Figure S11.** Over 180 processes were measured, matching pulse packets and single pulses with similar on time and different gap spacing between the pulses of the packet. This statistics of these datasets are shown for the deposition and ionization rates of pulse packets, normalized to their single pulse equivalents. A value higher than 1 indicates higher deposition/ion rates for the pulse packet. This is shown for in **a.** for pulse packets that had gaps <100 µs with pulse packets outperforming single pulses by a marginal amount at low off-times, before being overtaken by single pulse at higher on-times. In panel **b.** the same dataset is shown but against the gap spacing of the pulse packets, showing the significant decline in deposition and ionization rates as the gaps grow. These processes were done at a working pressure of 1.0 Pa, 2.2 $Wcm^{-2}$.

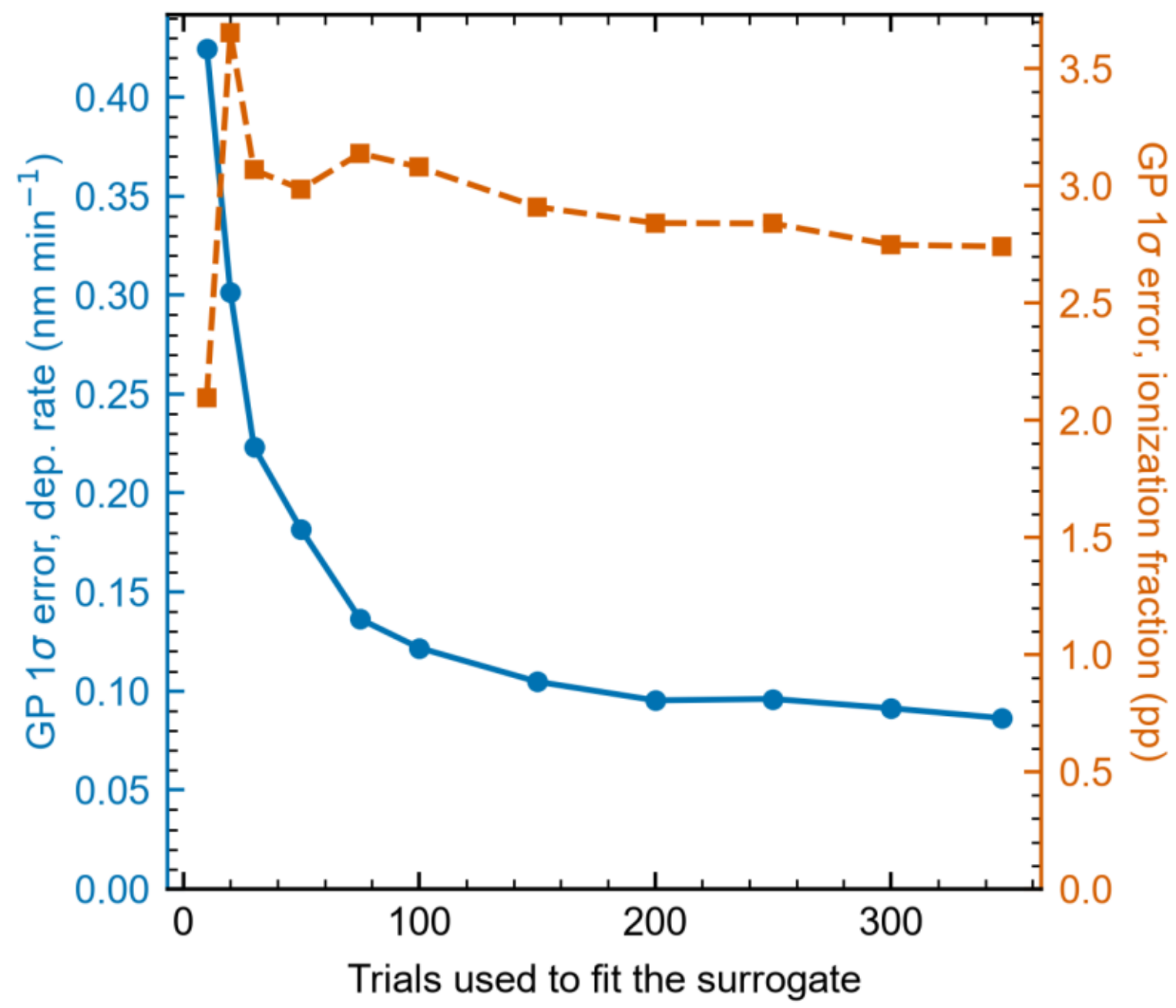

**Figure S12.** Surrogate uncertainty against sample number for the exploration campaign at 1.0 Pa / 2.2 $Wcm^{-2}$. A Gaussian process regressor was used as the surrogate, described in the *Experimental Methods* section. The curve represents the mean predictive standard deviation over 3000 random recipes, fit inside the campaigns lower and upper bounds. Uncertainty minimizes within the first 150–200 samples measured in the campaign.

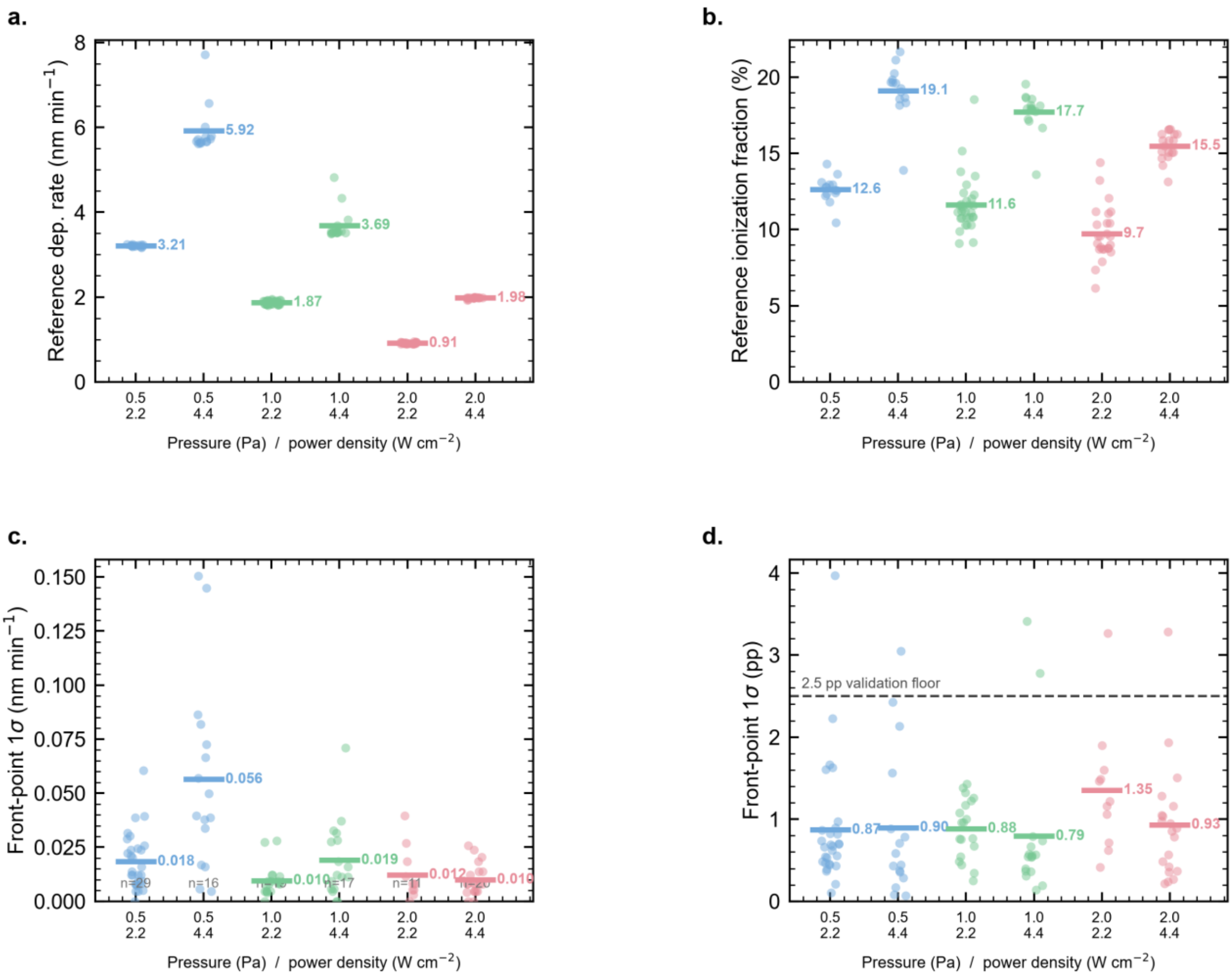

**Figure S13.** All campaigns run in this work included a drift reference, which was an identical process measured every 20 trials to check how deposition rates and ionization rates may be drifting on the target over time. These drift references are shown for deposition rate in **a.** and for ionization rates in **b.** The solid bar is the mean of the drift references, shown for all six campaigns done at working pressures of 0.5, 1.0, and 2.0 Pa at power densities of 2.2 and 4.4 Wcm$^{-2}$. Points along the Pareto front were then 'validated' as described in the *Experimental Methods* section, by collecting at least three measurements of the same process and checking their standard deviations against a floor. This validation statistics is shown for deposition rate in panel **a.** and for ionization rates in panel **b.** where the 2.5% limit for validated Pareto front points is shown. This indicates that points with 1 standard deviation above 2.5% were not validated and removed the front.

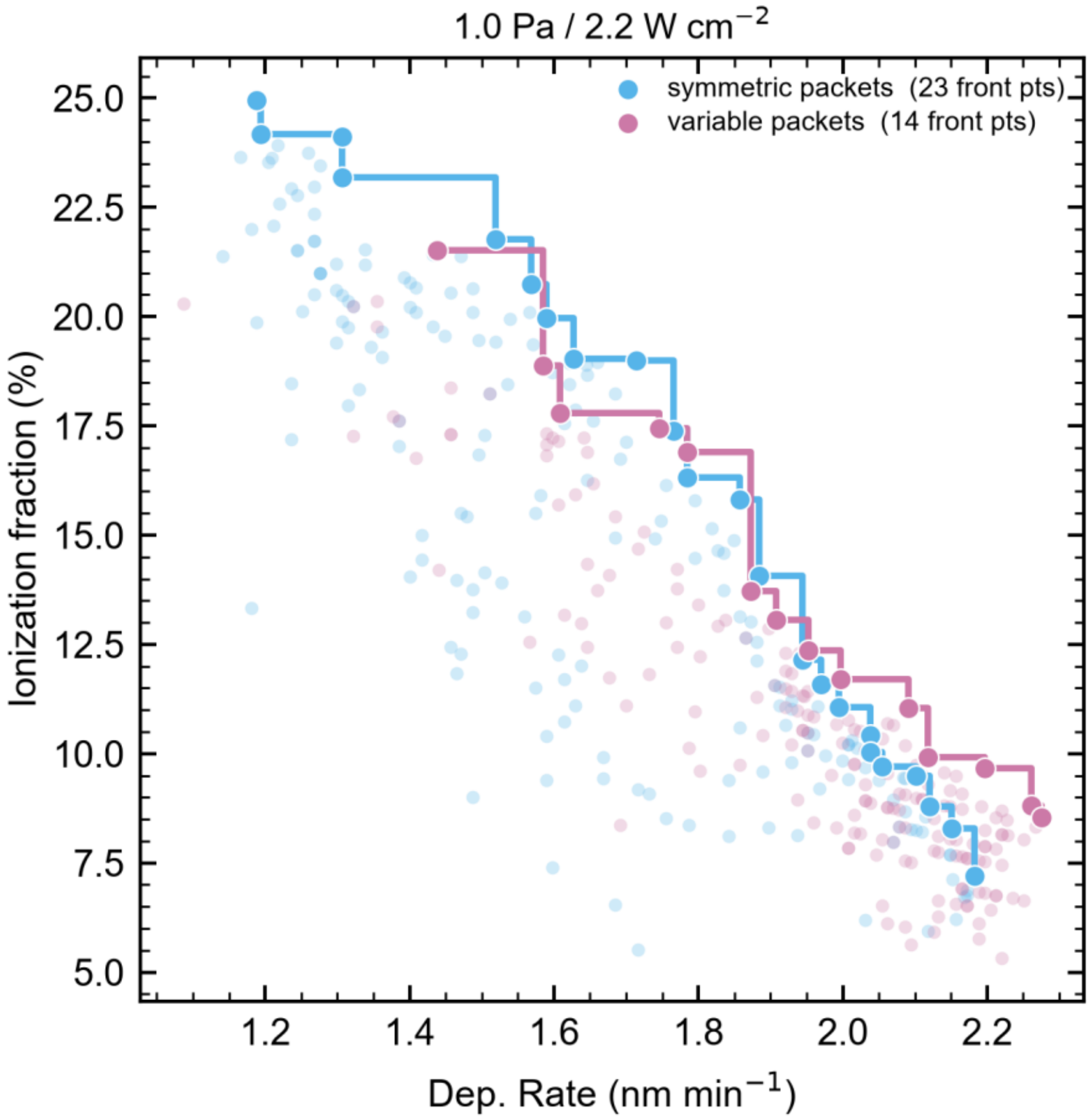

**Figure S14.** The differences between symmetric pulse packets (intra-packet gaps and pulse lengths are the same) vs. variable pulse packets (pulse-widths and intra-packet gaps can vary) was evaluated for at a working pressure of 1.0 Pa and power density of 2.2 Wcm$^{-2}$. Little noticeable difference is seen between the two validated Pareto fronts, with possible small gains to be found at low ionization fractions, but gains appear minimal. This follows nicely the gas rarefaction hypothesis proposed in the main body of this work, which no amount of variable pulse packet design would be able to overcome.

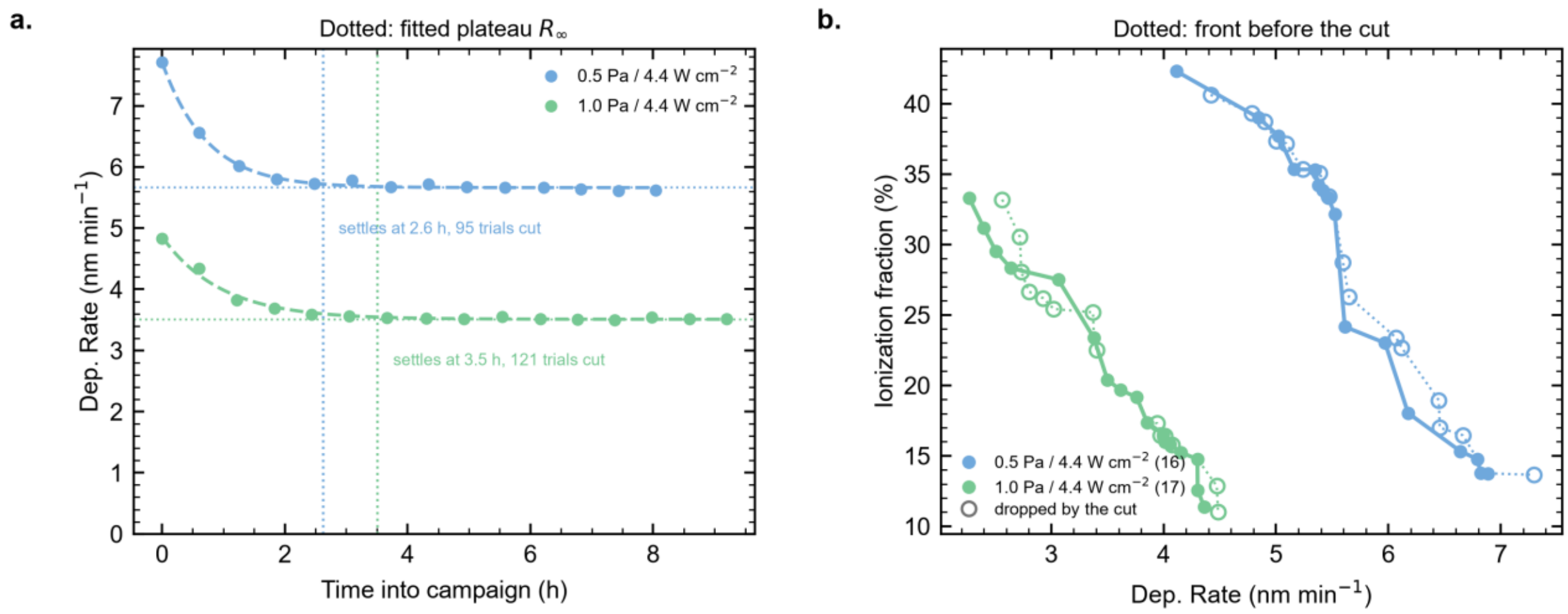

**Figure S15.** Drift measurements running during the Pareto front campaigns identified significant drift in the 1.0 and 2.0 Pa campaigns run at 4.4 Wcm$^{-2}$. This is shown in **a.** where ~2.6 – 3.5 hours are needed to stabilize the drift reference. When removing points in the campaign from those initial stages, the Pareto front shifts negligibly, shown in panel **b.** This error was mitigated by the validation procedure which always validates front-facing points at the end of the campaign.